\documentstyle[psfig]{mn}
\newif\ifAMStwofonts

\def\eso36{\hbox{ESO~3.6-m}}                          
\def\nin{\noindent}

\title[New ISOCAM Reduction Technique in the ELAIS Southern Region]{A New Method for ISOCAM Data Reduction -- I.
Application to the European Large Area ISO Survey Southern Field: Method and Results}
\author[C. Lari, F. Pozzi, C. Gruppioni et al.]
{\parbox[]{6.5in} {C. Lari$^{1}$, F. Pozzi$^{1,2,3}$, 
C. Gruppioni$^{4,3}$\thanks{e-mail: gruppioni@pd.astro.it}, H. Aussel$^{4,5}$, 
P.  Ciliegi$^{3}$, L. Danese$^{6}$, A. Franceschini$^{7}$, S. Oliver$^{8,9}$, 
M. Rowan--Robinson$^{9}$ and S. Serjeant$^{9}$}\\
\\
$^{1}$ Istituto di Radioastronomia del CNR, via Gobetti 101, I--40129 
Bologna, Italy\\
$^{2}$ Dipartimento di Astronomia, Universit\`a di Bologna, 
viale Berti Pichat 6, I--40127 Bologna, Italy \\
$^{3}$ Osservatorio Astronomico di Bologna, via Ranzani 1, I--40127 Bologna, Italy\\
$^{4}$ Osservatorio Astronomico di Padova, vicolo dell'Osservatorio 5,
I--35122 Padova, Italy\\
$^{5}$ Institute For Astronomy, 2680 Woodlawn Drive, 96822 Honolulu, Hawaii\\
$^{6}$ SISSA, via Beirut 4, I--34014 Trieste, Italy\\ 
$^{7}$ Dipartimento di Astronomia, Universit\`a di Padova, vicolo dell'Osservatorio 5, 
I--35122 Padova, Italy\\ 
$^{8}$ Astronomy Centre, CPES, University of Sussex, Falmer, Brighton BN1 9QJ, U.K.\\
$^{9}$ Imperial College of Science, Technology and Medicine, Prince
Consort Road, London SW7 2BZ, U.K.}

\date{Accepted 2001 March 15. Received 2001 February 21; in original form 2000 November 13}

\pagerange{\pageref{firstpage}--\pageref{lastpage}}

\pubyear{2001}

\def\LaTeX{L\kern-.36em\raise.3ex\hbox{a}\kern-.15em
  T\kern-.1667em\lower.7ex\hbox{E}\kern-.125emX}

\begin{document}

\label{firstpage}

\maketitle                  
                          
\begin{abstract}
We have developed a new data reduction technique for ISOCAM LW data and have
applied it to the European Large Area ISO Survey (ELAIS) LW3 (15 $\mu$m) observations in the southern 
hemisphere (S1). This method, known as {\it LARI} technique and based on the assumption of the 
existence of two different time scales in ISOCAM transients (accounting either for fast or slow 
detector response), was particularly designed for the detection of faint sources. 
In the ELAIS S1 field we obtained a catalogue of 462 15 $\mu$m sources
with signal-to-noise ratio $\geq$ 5 and flux densities in the range $0.45 - 150$ mJy (filling the
whole flux range between the Deep ISOCAM Surveys and the IRAS Faint Source Survey). The 
completeness at different 
flux levels and the photometric accuracy of this catalogue have been tested with simulations.
Here we present a detailed description of the method and discuss the results obtained by its
application to the S1 LW3 data.
 
\end{abstract}

                                                                                             
\section{Introduction}
The Infrared Space Observatory (ISO; Kessler et al. 1996) was the successor to the Infrared 
Astronomical Satellite (IRAS). ISO, besides carrying out detailed studies of individual 
objects and small regions, has provided an opportunity to perform survey
work at sensitivities of several orders of magnitude better than its precursor.
Thus, a significant fraction of the mission time was spent on field surveys.
The largest survey conducted with ISO is the European Large Area ISO Survey (ELAIS),
which provides a link between the IRAS survey and the deeper ISO surveys.
ELAIS is a collaboration between 20 European
institutes which involves a deep, wide-angle survey at high Galactic
latitudes, at wavelengths of 6.7 $\mu$m, 15 $\mu$m, 90 $\mu$m and 175 $\mu$m
with ISO (see Oliver et al. 1997 and Oliver et al. 2000 for a detailed description 
of the Survey). In particular, the 15 $\mu$m
survey was carried out with the ISO-CAM camera (Cesarsky et al. 1996) over
a total area of $\sim$ 13 deg$^{2}$, divided into 4 main fields and several
smaller areas. One of the main fields, S1, and one of the smaller areas, S2, 
are located in the southern hemisphere. S1 is centered at
$\alpha$(2000) = 00$^h$ 34$^m$ 44.4$^s$, $\delta$(2000) = -43$^{\circ}$
28$^{\prime}$ 12$^{\prime \prime}$ and covers an area of $2^{\circ} 
\times 2^{\circ}$, while S2 is centered at $\alpha$(2000) = 05$^h$
02$^m$ 24.5$^s$, $\delta$(2000) = -30$^{\circ}$ 36$^{\prime}$ 00$^{\prime \prime}$
and covers an area of $21^{\prime} \times 21^{\prime}$.
The whole S1 and S2 areas have been surveyed in the radio (at 1.4 GHz, Gruppioni et al. 1999; Gruppioni, Ciliegi, Oliver et al. 2000 in preparation)
in several optical bands and in the near-infrared (La Franca et al. 2000 in prep.; 
Heraudeau et al. 2000, in prep.). 

Since ELAIS is the largest survey performed by ISO and covers just the gap 
in flux density that exists between the IRAS Survey and the ISOCAM Deep and Ultra-Deep
Surveys (Elbaz et al. 1999), it was extremely important to obtain the best and most reliable 
possible results from these data through an accurate data reduction.

To this purpose, we have developed a new ISOCAM data reduction technique (the {\it LARI
technique}) especially
designed for the detection of faint sources. This method, designed by C. Lari
and based on the assumption of the existence of two different time scales in ISOCAM transients,
has been tested on ISOCAM-HDF data, providing excellent results in agreement
with those obtained with the {\it PRETI technique} (Starck et al. 1999).

Before attempting to reduce the entire ELAIS survey, we decided to apply the 
{\it LARI technique} to one single field, in order to test the capabilities
of our method and to adapt some of its tasks for this specific set of data. 
In particular, we have applied the
{\it LARI technique} to the 15 $\mu$m data in the southern ELAIS field S1, where 
most of the available multi-wavelength follow-up observations are available.
Here we present the results of the {\it LARI method} in S1, as well as the complete 
15 $\mu$m source catalogue obtained with this technique.

The paper is structured as follows:
in section 2 we present the survey strategy and parameters;
in section 3 we give a detailed description of the new data reduction 
technique that we have developed and used; in section 4 we describe the 
reduction and analysis of our data; in section 5 we present the results of tests 
made on simulated data; in sections 6, 7 and 8 we discuss the source photometry
the calibration accuracy and the astrometric corrections respectively, while in 
section 9 we describe our source catalogue and in section 10 we present our conclusions. 

The source counts obtained from these data will be presented and discussed in a companion
paper (Gruppioni, Lari, Pozzi et al. 2000, Paper II).

\section{The ELAIS survey observation strategy}
The S1 field, as well as the other ELAIS survey areas, was selected for its high Ecliptic
latitude ($| \beta | > 40^{\circ}$, to reduce the impact of Zodiacal dust emission), for its 
low cirrus emission ($I_{100 \mu m} < 1.5$ MJy/sr) and for the absence of any bright IRAS 
12 $\mu$m sources ($S_{12 \mu m} > 0.6$ Jy). 
In figure 1 the location of the S1 survey field is shown, overlaid on a Cirrus map
(the COBE normalized IRAS maps of Schlegel et al. 1998). IRAS sources with 12 $\mu$m
fluxes brighter than 0.6 Jy are also plotted.

\begin{figure}
\psfig{figure=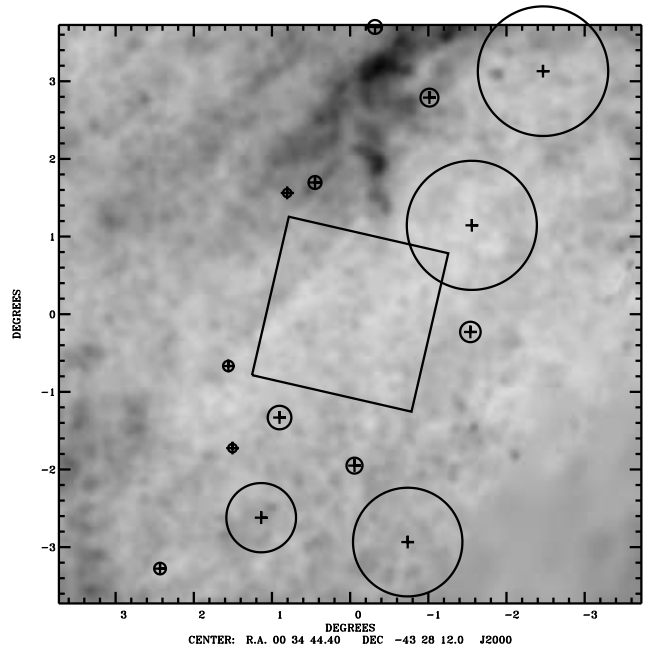,width=13.5cm,bbllx=50pt,bblly=0pt,bburx=554pt,bbury=360pt}
\caption{The sky position and orientation of the ISO S1 survey region overlaid on the COBE
normalized IRAS maps of Schlegel et al. (1998). The rectangle delimiting the S1 area is 
$2^{\circ} \times 2^{\circ}$. IRAS sources with 12 $\mu$m flux brighter than 0.6 Jy are also
plotted, with radius proportional to flux. The maximum 100 $\mu$m intensity shown (black) is 
1.5 MJy/sr.}
\label{S1_location}
\end{figure}

The ELAIS ISOCAM survey was conducted in raster mode with the LW2 (6.7 $\mu$m) and LW3
(15 $\mu$m) filters. The ISOCAM detector was stepped across the sky in a grid pattern,
with about half detector width steps in one direction and the whole detector width steps
in the other. In this way, the reliability was improved as each sky position was observed 
twice in successive pointings and the overheads were reduced because each raster covered a 
relatively large area (40$^{\prime} \times 40^{\prime}$). At each raster pointing (i.e.
grid position of the raster) the 32$\times$32 ISOCAM detector was read out several times. 
Table 1 describes the observation parameters for the LW3 filter.

\begin{table}
\centering
  \caption{S1 LW3 observation parameters}
  \label{parmstab}

\begin{tabular}{lc}
 &  \\ \hline
Parameter & LW3 (15 $\mu$m) \\
 \hline
Band width & 6 $\mu$m \\
Detector Gain & 2 \\
Integration time & 2 s \\ 
Number of exposures per pointing & 10 \\
Additional number of exposures to stabilise & 80 \\
Pixel field of view & 6$^{\prime \prime}$ \\
Number of pixels & 32 $\times$ 32 \\
Number of horizontal and vertical steps & 28, 14 \\
Step sizes & 90$^{\prime \prime}$, 180$^{\prime \prime}$ \\ 
total area covered & 3.96 deg$^2$ \\
 \hline

\end{tabular}
\end{table}

\section{Lari Technique: Generalities}
\label{laritecnique}

As already described in detail by Starck et al (1998),
ISOCAM data obtained with the long wavelength detector (LW) are affected by several problems. 
The two main effects, which become more important the deeper we push for source detection, 
are produced by cosmic ray impacts (`glitches') and transient behaviour (slow response of 
the detector to flux variations). 

Usually, `glitches' can be divided into three categories: {\it common}, {\it faders} and
{\it dippers}, according to their behaviour, decay time and influence on the pixel responsivity.
Slow decreases of the signal following cosmic ray impacts are called {\it faders}, while 
depletions in the detector gain, followed by a reduction of the pixel sensitivity very slowly 
recovering afterwards (see figure \ref{pixel_example}, $top$ panel), are called {\it dippers}. 
These two effects are believed to be associated
with proton or $\alpha$ particle impacts on the detector, while cosmic ray electrons
produce {\it common} `glitches'. {\it Common} glitches only last one readout and their
decay time is relatively fast (lasting only a few readouts), while {\it faders} and {\it dippers} 
have much longer lasting impact
on the pixel sensitivities. So, the number of frames affected by the latter is much higher than 
in the case of {\it common} glitches, the sensitivity of pixels taking from tens to hundreds of 
seconds to recover completely. However, {\it common} glitches are much more frequent than
{\it faders} and {\it dippers} and, if not correctly removed, they may look like sources
on the maps and produce false detections. For this reason, the data cleaning is
an extremely delicate process, which requires great care in order to produce highly
reliable final maps and source lists.
 
The {\it LARI method} was mainly developed to overcome the main problems affecting ISOCAM LW
data and to give better quality maps and as complete and reliable source catalogues as possible.
Analogously to the {\it PRETI method} (Aussel et al. 1999), our algorithm corrects the cube of 
ISOCAM data for cosmic rays and transient effects before reconstructing the images and 
carrying out source detection.  

The model on which the {\it LARI method} is based (described in detail in Appendix A), rests 
on the assumption that
the incoming flux of charged particles generates transient behaviour producing 
two different time scale effects: a fast ({\bf breve}) and a slow ({\bf lunga}) one.
The latter component accounts for the slow response of the detector and is essential in
recovering the transient effects of the {\it dippers}.
Each of the two time scales is associated with an independent reservoir of charge, which 
decays with this characteristic time-scale towards the contacts 
(i.e. a multi-component model for semiconductors).
These two reservoirs of charge are fed by both incident infrared photons 
and cosmic rays. The latter are also able to trigger a fast charge release   
towards the contacts (`glitches'). 
When a cosmic ray particle hits the detector, the quickly-varying charge reservoir {\bf breve} 
is on average increased, while the slowly decaying charge reservoir is quickly forced to 
release part of its charge content. 
Thus, while the {\bf breve} component is fed by a large fraction of the incident photons 
(around 40-45 \%), the {\bf lunga} one is fed only by a few percent of them. The remaining 
fraction is very quickly forced towards the contacts ({\bf prompt} component). Due to 
differences between the two time scales of about a factor of 20 when the process reaches 
stabilisation, the {\bf lunga} component collects a higher total amount of charge with respect
to the {\bf breve} one.

The value of both time constants depends on the signal level 
(which is fixed by observations) such that the lower is the signal, the larger is
the time constant.
Our model simply assumes the time constant to be inversely proportional to 
the amount of accumulated charge.

To first order, {\it faders} are described in this model as discontinuities mainly in the {\bf breve} 
charge reservoir, caused by the cosmic ray impact. 
Similarly, to first order the {\it dippers} are discontinuities mainly in the {\bf lunga}
charge reservoir.
The maximum depth of the {\it dippers} is determined by the fraction of the flux feeding the 
{\bf lunga} reservoir. The overwhelmingly large majority of pixels are well-fit by a {\bf lunga} 
fraction of $\sim 0.1$, implying 
that {\it dippers} cannot exceed one-tenth of the sky background level. Very occasionally however 
some dippers exceed this threshold. To account for these, an additional
zero-point dark current `offset' can be set, so the maximum depth is not larger than one-tenth of the 
revised total background. Incidentally, the presence of dippers in the dark current records (that have 
zero background) shows that this `offset' is a general property of the detector, almost certainly fed
by the thermal noise.

\section{Application to ISOCAM LW ELAIS data}
The application of our model to the ISOCAM LW data obtained in the ELAIS fields
required some particular adaptation of the algorithms and the construction of
some `ad hoc' procedures necessary to overcome the specific problems generated by
the chosen observational strategy. 

The main cause of problems in the ELAIS data is the very short integration time. In fact, 
the time spent by the detector on each readout in these observations is only 2 seconds 
(see Table \ref{parmstab}) and the total time spent on each raster pointing is 10$\times$ 
the integration time. The short observing time over each raster position (10$\times$2 sec) 
not only affects the signal-to-noise ratio, but it has two major negative effects on the data:
\begin{itemize}
\item[1)] since it is shorter than the fast time scale, it makes strong glitches 
hide real sources; 
\item[2)] only a  fraction (60\%) of the total incoming flux is recorded during
each exposure, thus causing large photometric errors.
\end{itemize}

In our model, glitches are treated as discontinuities in the charge ($Q$) reservoir,
with constant parameters $a$ and $e$ (see Appendix A). However, immediately after the maximum of a glitch,
the detector is considered to behave normally under a constant (over the raster pointing)
flux $I$. This is not completely correct for very strong glitches, which may cause the signal
immediately following the maximum to be higher than predicted and this is mostly true
for short integration time observations (i.e. 2 seconds like ELAIS).

Moreover, the relation between the increment of the {\bf breve} component and the decrement of
the {\bf lunga} one is not constant. In fact, cosmic rays producing a higher increment 
in the {\bf breve} reservoir than in the {\bf lunga} one look like {\it faders}, while
in the opposite case we have {\it dippers}. In the ELAIS data there are {\it dippers} without an 
initial glitch spike: generally we find the glitch feature in a contiguous pixel but, 
very rarely, it is completely absent. 

Another problem arises from the fact that the Point Spread Function (PSF) is spatially
under-sampled in all ISOCAM LW3 observations with pixel-field-of-view (PFOV) lens greater than 1.5 
arcsec (in our case PFOV = 6 arcsec). Thus, any position determination method applied to individual 
point sources gives biased results for under-sampled data (the worse sampled the data, the more 
the resulting position is centered on a pixel).
This bias can be corrected to some extent and the source position can be improved up to a fraction
of the pixel size by taking into account the {\it a priori} raster pattern.
In any case, the PSF is not unique for all sources, but it depends strongly
on the source location within a pixel. This PSF, corresponding to the position of a source in 
the raster map and with an average FWHM of $\simeq$10$^{``}$, will be referred as 
{\bf ``effective PSF''} throughout the paper. Moreover, in the ELAIS data each pixel
in the final raster map comes from the combination and projection on the sky
of different overlapping single images. For this reason, a source in the raster
map is produced by the combination of different source images, where the source
has different flux distributions, depending on its location within the pixel
of the single images. This is a serious problem which affects source 
detection, source photometry and the completeness of the catalogue.
In this work, we have carefully analysed and tried to quantify this 
combination effect by performing simulations (see section \ref{simulations}). 

\subsection{ELAIS Data Reduction with the Lari Method} 
\label{data_reduction}
\begin{figure*}
\begin{minipage}{180mm}
\centerline{\psfig{figure=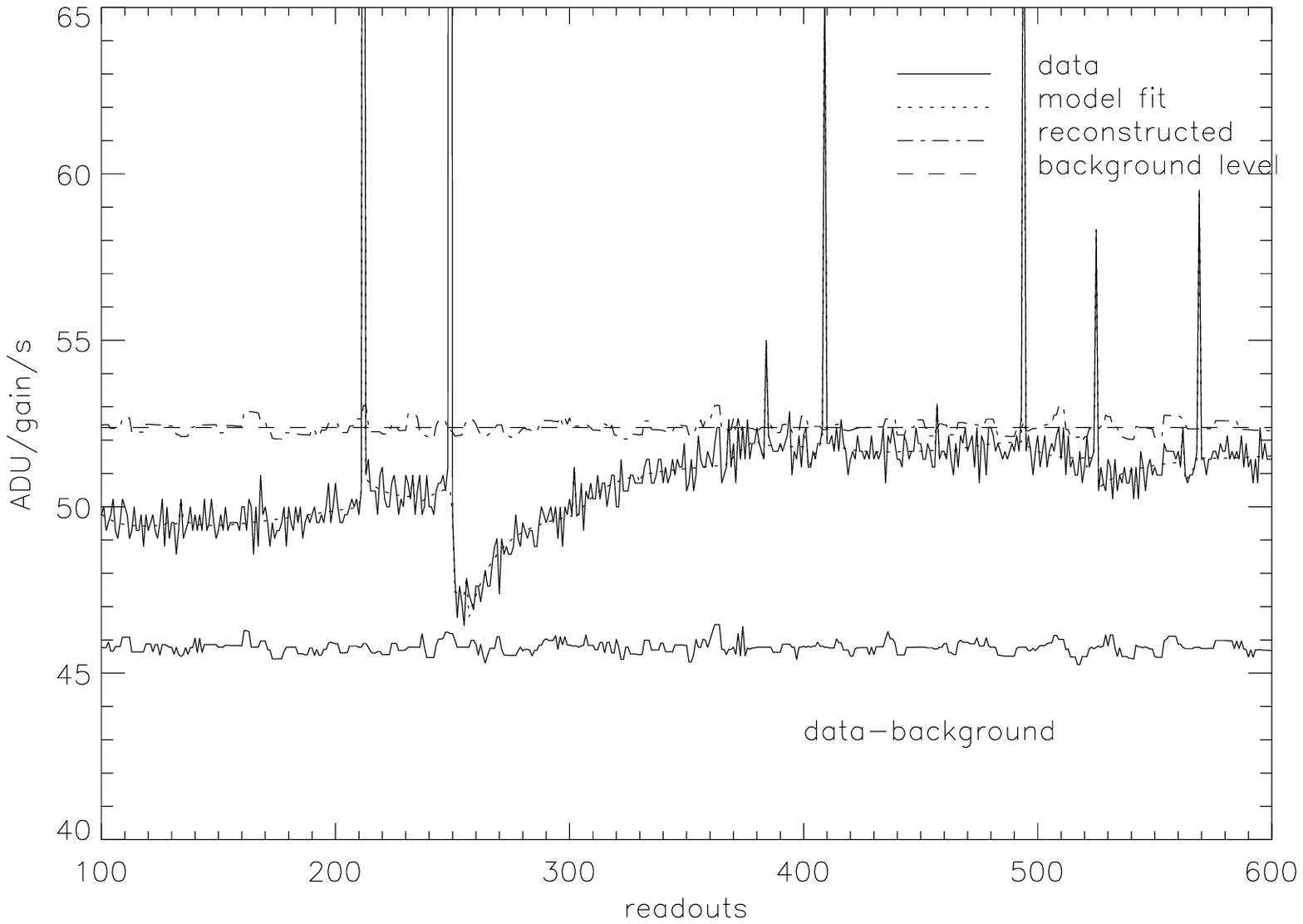,width=14cm}}
\centerline{\psfig{figure=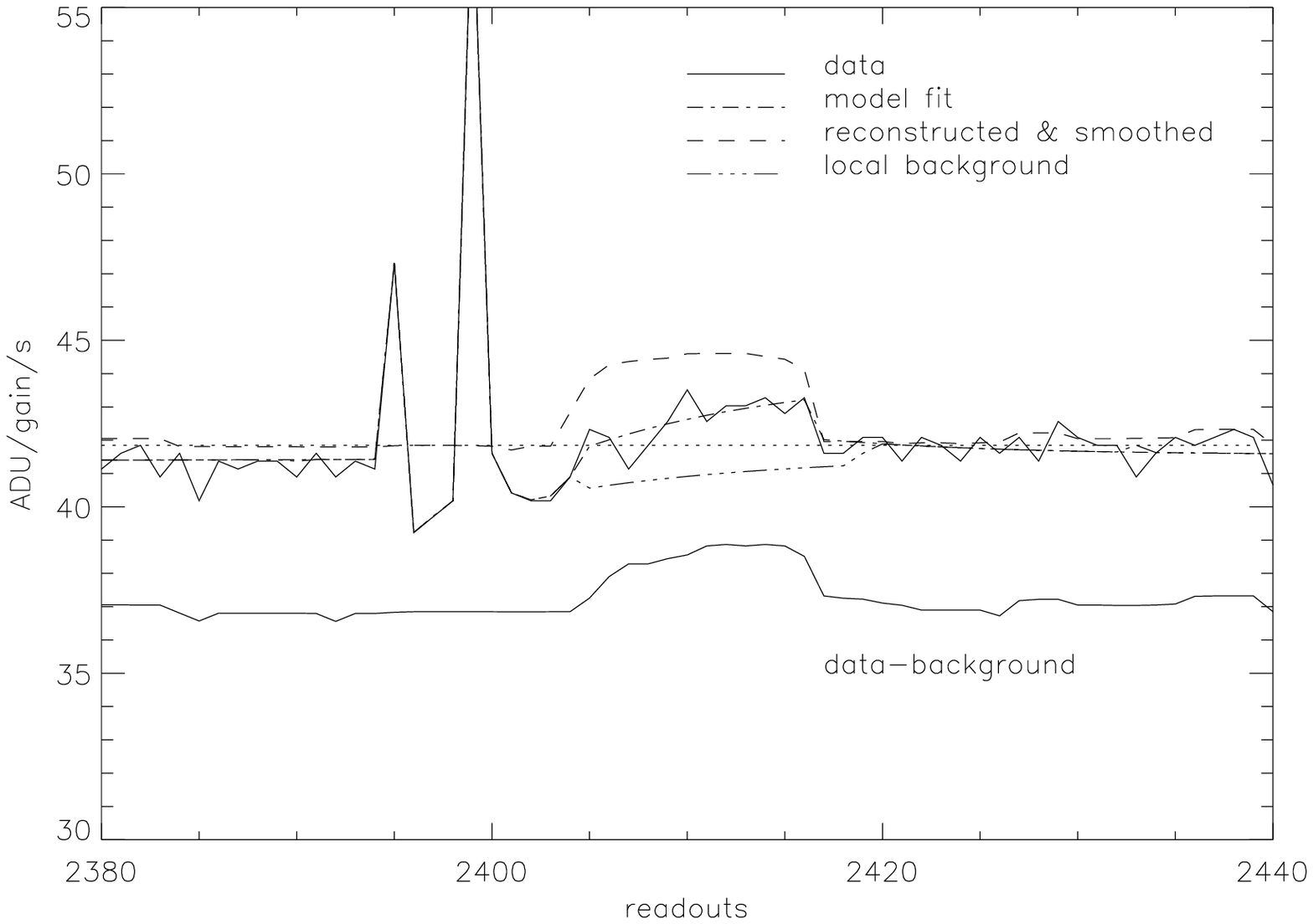,width=14cm}}
\caption{Example of real and model data through the pixel history. The 
solid line represents the data and the dot-dashed line the best-fit model, 
while the dashed line is the data corrected for transients and deglitched.
The dotted horizontal line is the assumed background level. 
In the $top$ panel the characteristic ISOCAM LW transient behaviours due to 
cosmic rays are shown. The raw data are clearly affected by many `glitches', as well as by 
strong `dippers' and `faders', and by upward transients. The $bottom$ panel is a zoom in the history
of a pixel, which shows how our model works in fitting the data and reconstructing the 
stabilized signal when the pixel sees a source.}
\label{pixel_example}
\end{minipage}
\end{figure*}

All the codes developed for data reduction with the {\it LARI method} are written
in the Interactive Data Language (IDL) and the whole data reduction and analysis
has been performed with IDL software.
The main process of ISOCAM LW data reduction with the {\it LARI method} consists
of several basic steps. First, the raw data are
converted into a raster structure containing all the information about the observation
(pointings, instrument configuration, etc.) as well as the single images (one for each 
pointing, which are combined together to give the final raster image). 
The images are then converted to ADU/gain/s and the dark current subtracted. 
These first two steps are performed using the CAM Interactive Analysis (CIA) package. 
Next, the data are corrected for short time cosmic rays and the affected
readouts are masked before copying the ``de-glitched'' data to a new structure 
(called ``liscio''). This structure contains also the initial guess for the parameters defining the 
{\bf lunga} and {\bf breve} reservoirs of charge
and information about the main `glitches' and `dippers' (derived by the deglitching process).
The task which performs the first guess for parameters, also evaluates 
the background and the minimum `offset' to be added to the data in order to have the `dipper' 
depth in the range allowed by the model (one tenth of the sky background, see section \ref{laritecnique}). 
This is done for each pixel.
The code not only finds the stabilization background level, that is, the zero level for data fully
recovered from transients, but models the `glitches', the sources and the background with all 
the transients over the whole pixel history. Moreover, our code is able to predict the trend we 
would have on each raster position if only the stabilization background flux was hitting the 
detector (starting from the previously accumulated charges, i.e. the `local background').
The excess with respect to this `local background' represents the flux excess not recovered from
transients. The maps of this excess, after flat fielding,
are called `{\bf unreconstructed}' maps and (in case of a good enough fit) represent the effective flux collected 
by the detector during the raster exposure. In this paper, fluxes obtained from `{\bf unreconstructed}'
maps will be named {\it fs}, while fluxes measured on `{\bf reconstructed}' maps (i.e. reconstructed
from transient effects) will be named {\it fsr}. These two fluxes that we can measure for a source
are shown in figure \ref{pixel_example}($bottom$ panel): the dot-dashed line
represents the `{\bf unreconstructed}' data, while the dashed line represents the data `{\bf
reconstructed}' for transients.
   
With our code,  we created a model data-set for the deglitched data, reproducing 
not only the source signal, but also all the transient effects affecting the data. In figure 
\ref{pixel_example} an example of pixel history is shown, together with the background and 
data models obtained with our algorithm.

In outline, the fitting algorithm starts with the brightest glitches in the raster, assumes 
discontinuities at these positions, and tries to find a fit to the time-lines that satisfies
the solid-state physics of the detector. If no acceptable fit is found, the next fainter glitch 
is considered as a potential discontinuity, and so on. 
Because of the reduced number of useful readouts in the ELAIS raster data, in the fit we use fixed
default values for all the pixels (the physical parameters scaled only for the background level),
leaving as free parameters only the charge values at the beginning of the observations and at the 
top of glitches.
  
By successive iterations, the parameters and the background for each pixel are adjusted to 
fit the data better, until the rms of the difference between model and real data is smaller 
than a given amount (e.g. 0.2 ADU/gain/s ). Note also that the effects generated by the 
presence of glitches in the nearby pixels are considered by the fitting algorithm.
The code recognizes sources above a given flux level, which decreases as the reduction improves
the fit. In the pixels around relatively strong sources ($>$ 1.3 ADU/gain/s) we force the fit to
find sources, leaving the fit level free. 
Once a satisfying fit is obtained for all the pixels over the whole pixel history, the flat field 
is computed from the stabilization level of the background. In the raster structure
we set the flat fielded smoothed differences of: a) data readouts minus local fitted
background (`unreconstructed data'); b) fully recovered intensities minus stabilization 
level (`reconstructed data'). Glitches and bad data are masked and this mask is stored in the raster 
structure to be used later in the map creation.

Then the reduced images per raster pointing are computed and corrected for flat-field
distortions. 
Finally, the images are combined together to create the final raster maps (one for each raster 
position), where we then look for source detection.

A final reduction stage is performed after source extraction, simulating the data we would 
have from these detections and correcting the pixel fit, forcing the algorithm to recognize the
source whenever this had not happened correctly (i.e. the source had been recognized only in one 
of the two overlapping
single pointing images and not in the other).

We will now enter into more detail on the map creation and 
source detection processes.

\subsection{Map creation}
\label{mapcreation}
Once the images, corresponding to each raster position, have been created by averaging 
together all the time-scans relative to that pointing, they are converted from ADU/gain/s
to mJy/pixel using the ISOCAM User's Manual calibration factor (e.g. dividing by 1.96)
and flat fielded.
Also the number of un-masked times-scans (raster.npix) are scaled with flat-field coefficients,
on the simple assumption of a constant noise in the data prior to flat-fielding.   
 
After that, they are projected onto a sky map (raster image) using a simple TAN projection. 
The algorithm used is part of the CIA package ({\it projette.pro}). 
It computes the values of pixels on the sky map by averaging the pixel values in the 
single images, with a weight equal to the number of useful time-scans, to give the 
pointing image (raster.npix). 
For each raster map, two corresponding maps are also constructed, with the same
size and sky orientation. One is the map containing, for each pixel, the number
of frames coadded together (excluding the masked pixels) to obtain 
that pixel value in the raster (``NPIX'' map). The other is the ``RMS'' map,
where the rms of each pixel has been computed by scaling the measured mean rms of the 
central part of the map according to the inverse square root of ``NPIX''.

In projecting each raster pointing onto the sky, the algorithm takes
into account the field distortions of ISOCAM, as measured by Aussel et al. (1999).
These distortions are a chromatic effect which cause the pixel size to be non-uniform
on the sky (border pixels are larger in area than central pixels). This effect must
be considered also when computing the flat-field. For a more
detailed description of the ISOCAM field distortions see Okumura 
(2000).

\begin{figure*}
\begin{minipage}{170mm}
\centerline{\psfig{figure=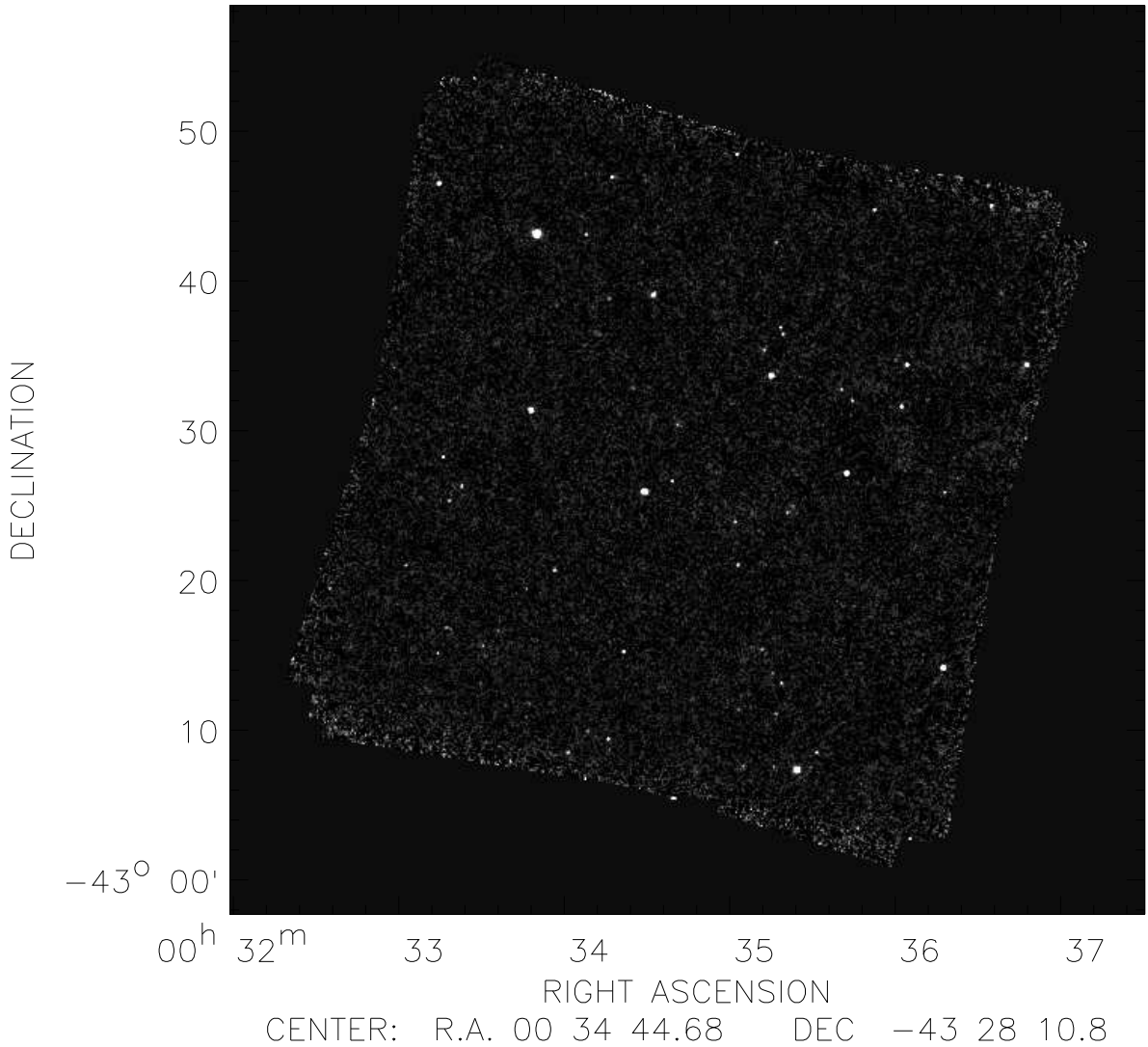,width=22cm,bbllx=4pt,bblly=360pt,bburx=508pt,bbury=720pt}}
\caption{Grey-scale image of the central raster $S1\_5$. This map derives from the
combination of three single observations and the higher noise
level at the borders, where the overlap between the three pointings is not
perfect, is clearly visible.} 
\label{s1_5_map}
\end{minipage}
\end{figure*}

In figure \ref{s1_5_map} the grey-scale image of the central raster map (S1$\_$5) is shown.
Note that this image, $S1\_5$, is the combination of three single observations, 
this field having been observed with a redundancy of 3 compared to the other rasters.

The projection algorithm strongly affects the appearance of point sources on the map, having
the general effect of smoothing the PSF over several pixels. As we will show in section
\ref{simulazione}, the peak flux values of sources with the same total flux can differ
significantly, by a factor of up to 2, depending on the source position within the raster pixel. 
 
\subsection{Source detection}
\label{source_ext}
Before performing the proper source detection (on the final maps) to produce our definitive
source catalogue, we have identified candidate sources inside the pixel histories. This
was also very useful to check the confidence level of our fit to the data. 
Since during data reduction we have created a model for the background, we identified
sources in the history of pixels from their flux excess above the background over the single 
time-scans. We inspected by eye every excess greater than 0.7 ADU/gain/s, correcting 
the few cases corresponding not to real sources, but to algorithm failures 
(by re-setting the parameters and starting again with model fitting until convergence is 
achieved). We have found 
that our method is very conservative, in the sense that cases where a spurious source is 
created by the algorithm constitute a very small fraction of the total number of correct 
detections, while the fraction of good sources missed by the fitting algorithm is rather 
significant (since sources are normally seen on several pixels, a lost detection on the pixel history 
does not necessarily mean the source is lost on the map).

Only faint sources remaining undetected on the map (because of these failures) contribute 
to real incompleteness, while for most of the brighter sources the failures result in a 
decrease of their total flux.
In the final stage of our reduction (see below), we re-project the sources detected on the 
raster map into the pixel time-line, allowing a better fit of the data for all the sources
that will appear above the interactive check threshold. This job significantly reduces
the flux defect for the detected sources (but is of course unable to recover sources
that fell below the detection threshold, i.e. to correct for incompleteness).
 
Concrete determination of the fraction of detected sources versus real sources, which leads 
to the estimate of completeness and reliability of our source catalogue, have been performed 
using simulations. This will be discussed in section \ref{simulations}. 


After this preliminary check, we have searched for detections in the single calibrated 
images by selecting and visually inspecting all pixels with flux $> 0.4$ mJy / pixel. 
In this case too, we have performed again the reduction
for those pixels where the algorithm had failed to fit the data, thus producing a false detection.  

These two checks on candidate sources, which required corrections and further cleaning for 
some pixels, provided very reliable (not complete!) source lists and images.
After that, we could be confident that almost no spurious sources were present in our
data-set. Therefore, we could proceed to the proper source detection.
We must now point out that all the checks performed on the single pointing
images and pixel histories do not guarantee that all these (and only these)
sources will then be detected on the final raster map. In fact, as already
discussed, the raster pixels are produced by combining together different
single pointing images, where the same source could be located in different
positions inside the pixels, thus being for example above the 0.4 mJy/pixel
check threshold in one image and below this threshold in the other. For this
reason, the list of sources obtained in the single pointing images for our
preliminary checks are not always coincident with the final list derived on 
the raster map, where each source is determined by different effects. Moreover,
in the raster map creation there are also distortion effects. Thus, the
completeness and reliability of our final source lists can be tested only
through simulations.

The source detection is done on the final raster maps, but is done using the signal-to-noise 
ratio. First, we have selected all pixels above a low flux threshold (0.1 mJy / pixel) using 
the IDL Astronomy Users Library (accessible via the World Wide Web home page 
http://idlastro.gsfc.nasa.gov/homepage.html) task called {\it find}. This algorithm
finds positive brightness perturbations (i.e. stars) in an image, returning centroids and 
shape parameters (roundness and sharpness). Then, we have extracted from the selected
list only those objects having a signal-to-noise ratio $\geq$ 5.

As discussed earlier, the {\it LARI method} is able to `reconstruct' the source flux 
from transient effects. However, as we will clearly show with simulations (see section
\ref{simulazione}), the algorithm does not `reconstruct' faint fluxes, corresponding
to sources that it is not able to recognize. Therefore, faint sources have {\bf `reconstructed'} 
fluxes similar to their {\bf `unreconstructed'} ones, while for bright and correctly
{\bf `reconstructed'} sources the {\bf `unreconstructed'} flux is, on average, about
1.7 times smaller than the corresponding {\bf `reconstructed'} flux. 
For this reason, to have a homogeneous flux determination for all our sources (both bright
and faint), we have chosen to run the detection algorithm on the {\bf `unreconstructed'} maps.
The correction for transients effects has then been performed individually in a second step, by using 
for each source its {\bf ``effective PSF''} when deriving its total flux (``auto-simulation''
procedure, see section \ref{autosimulazione}). 

In order to achieve better position determination, we have run the detection algorithm
on higher resolution maps, obtained by re-binning the original raster maps with a pixel
size of 2 arcsec.
The positions and fluxes given in output by {\it find} are determined by a convolution 
with a Gaussian PSF of 
given full width at half maximum (FWHM). We have chosen a FWHM = 9.8 arcsec, 
which is slightly smaller than the average FWHM of the ELAIS LW3 {\bf ``effective PSF''}.\\
The fluxes given by {\it find} are peak fluxes (mJy / pixel), which, coming from a 
Gaussian convolution,
do not always correspond to the real source peaks. Therefore, for source peak flux ({\it fs}) we have
given the maximum pixel value found within a box of 4 $\times$ 4 arcsec around the
maximum  given by {\it find}.  
To obtain the total fluxes (in mJy) we have used (and compared) two different methods: 
direct aperture photometry on the maps and ``auto-simulations'', as discussed 
in detail in section \ref{flux}.

\section{Simulations}
\label{simulations}

Because the raster maps on which we have performed the source detection are derived from the
combination of several single pointing images, the only way to evaluate the effects produced
on sources in the combined maps is through simulations.

With simulations, we can study the completeness and reliability of our detections at different 
flux levels and estimate the source positional accuracy and the internal calibration of the 
source photometry.

We added  randomly distributed point sources to each of the three overlapping central raster maps 
(S1$\_$5, S1$\_$5$\_$B and S1$\_$5$\_$C) at five different total fluxes 
(200 at 0.7, 150 at 1, 200 at 1.4, 150 at 2 and 150 at 3 mJy). It must be pointed out that
our simulations have not been performed over the entire flux range covered by our survey,
but only at the faint end. The reason is that we choose to sample with a high statistical
significance the flux range more affected by incompleteness effects due either to mapping
undersampling or data reduction method failures. 

In a similar way we get simulations on the combined mosaiced map for which we have 50 sources
at the same five flux levels.  

To perform the simulations, we have created a high resolution map ($1^{``}$) with 
simulated source, taking the ISOCAM PSF into account. The PSF has been 
successfully modeled by Okumura (1998) for stars. 
The PSF varies with the wavelegth, and since the ISOCAM filters are large, the 
shape of the PSF depends on the assumed spectrum of the point source. For our 
purpose, we have recomputed a model, following the prescriptions of Okumura 
(1998) but using a spectrum of the form $f_{\nu} = constant$, that is a closer match 
to the expected galaxy spectrum than the Rayleigh-Jeans form used for stellar 
spectra. The resulting PSF is larger than the one computed on stars.
 
Inverting the flat field and converting in ADU/gain/s we obtained the flux 
excess corresponding to the simulated sources. The value of this flux excess was then added to
the real pixel histories (containing glitches and noise) by using {\it Lari model}, to obtain 
the simulated data cube.
This simulated flux, included in the ``liscio structure'' (see section \ref{data_reduction}), 
have been reduced exactly in the same
way as we did for all the original data structures, doing the same checks and repairs. 
In the produced maps we extracted the simulated sources with the same procedures used for the 
real rasters, measuring the resulting positions and peak fluxes. 
The peak fluxes measured after the data reduction will be referred as {\it fs} and {\it fsr}
(as well as for real sources) respectively for {\bf `unreconstructed'} and {\bf `reconstructed'}
maps. The corresponding theoretical peak fluxes associated to the excess flux maps, not
reduced and containing neither glitches or noise, will be named {\it f0} and {\it f0r}. 
These two sets of parameters then allow to evaluate separately the effects produced by 
the ELAIS observational strategy and the ISOCAM instrument only ({\it f0} and {\it f0r}; 
see section \ref{autosimulazione}) from the effects 
produced by the {\it LARI reduction method} applied to ELAIS data ({\it fs} and {\it fsr};
see section \ref{simulazione}).

\subsection{Theoretical transient behaviour of the detector}
\label{autosimulazione}

By simulations of the theoretical transient behaviour of the detector, we mean simulations of the effects 
due to the finite spatial resolution (6 $^{\prime \prime}$) and to the finite integration time (10$\times$2 sec) of 
our observations. 
We need to consider the spatial resolution of our observations, since the PSF is comparable in size
to a pixel, causing the observed point source to depend strongly on its position on scales smaller than
the pixel size.

Regarding the finite integration time, we need to simulate the fact that 
the CAM detector does not reach immediately the level corresponding to a given input flux, but 
needs a certain time to stabilize (see section \ref{laritecnique}). This stabilization effect, which 
means that only a fraction of the incident flux is detected, is not constant, but depends on the length of the
time spent by the detector on target (not always the same) and on the amount of masking in a pixel
(due to `glitches' and uncertainties on the time spent on positions).
 
With the positions measured on simulated maps we can simulate how sources would appear in the ELAIS 
rasters if no noise and `glitches' were present.
To do this, we created two maps for each raster. The first is obtained by projecting 
back the simulated sources, injected on measured positions, onto the single pointing
images and then computing the resulting raster map.
The peak fluxes measured on it are the `reconstructed' peak fluxes: $f0r$.
For the second map we went back to the pixel history, predicting the behaviour due to the finite 
integration time transient (only theoretical, without any noise) and then producing a raster
map. The peak fluxes measured on it are the `unreconstructed' peak fluxes: $f0$.

\begin{figure}
\centerline{\psfig{figure=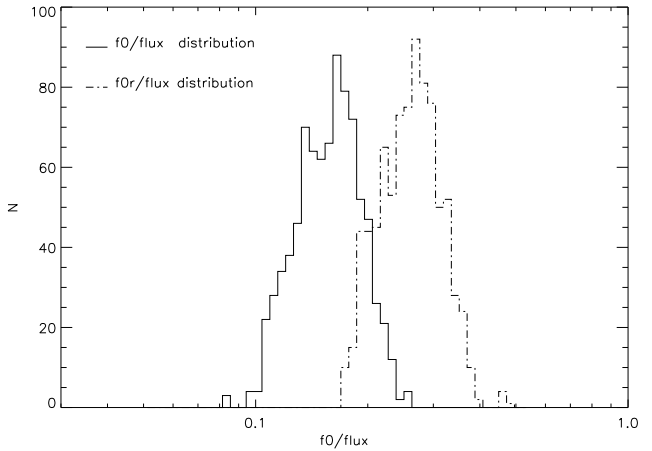,width=9cm}}
\caption{Distribution of normalized peak flux for mapping effects only, $f0r/flux$, and for
mapping $+$ source transient, $f0/flux$, for all the 850 simulated sources .}
\label{f0n_f0rn}
\end{figure}

Figure \ref{f0n_f0rn} shows the distribution of the peak fluxes 
(`unreconstructed' peak flux: $f0$; `reconstructed' peak flux: $f0r$) normalized to 1 mJy 
(divided by the corresponding input flux, i.e. the effective PSFs), for simulated sources. 
The effect of mapping is the main responsible for the large spread of values in the figure, 
because the distribution of the $f0r/f0$ ratios has a rms of only 0.09 (0.05 for the 
repeated field S1$\_5$) around a mean value of 1.66. The position uncertainty is only a 
minor contributor to the observed spread, since it causes only $\sim$14\% of the $f0$ and 
$f0r$ distribution dispersion.

The ratios of sources detected over the total number of injected sources, due to PSF 
under-sampling and finite integration time are reported in the second column of Table 
\ref{tab_compl} for each input flux.

The simulation of the mapping and mapping $+$ transient effects provides the estimate of
the individual PSF for each source and gives a technique to be used to derive total fluxes
for all the sources: for each source we will have two individual PSFs, one from $f0r$ and
one from $f0$.

As we will show in section \ref{flux}, there is a tight correlation between the peak flux
of a source and its theoretical peak flux due to mapping and transients only (``effective
PSF''). These 
``effective PSFs'' can be used for aperture flux determination (using small radii) and we will show that 
the total fluxes derived from the observed peak fluxes correspond very well to the 
aperture photometry ones.

\subsection{Real transient behaviour of the detector}
\label{simulazione}

Since the data reduction method can cause incompleteness in the final source list, we 
must take into account the effects produced by the fit when deriving the corrections to
be applied to our catalogue. In fact, our data reduction method is based on a fitting algorithm and, 
depending on how well this is able to model the background, the `glitches' and the sources,
our catalogue will be more or less complete. For this reason, with simulations we have also
tested the effects of {\it Lari model} on the final data products.

As stated above, to test the data reduction method, we have followed the 
same procedure for the simulated data that we used for real data. These simulated data cubes 
contain both real sources
and simulated ones. They have also the same rms noise, all the `glitches', `faders', `dippers' and 
background transients as the original data. The confusion will be slightly increased, but this
effect is not critical for ELAIS data. 

\begin{figure}
\centerline{\psfig{figure=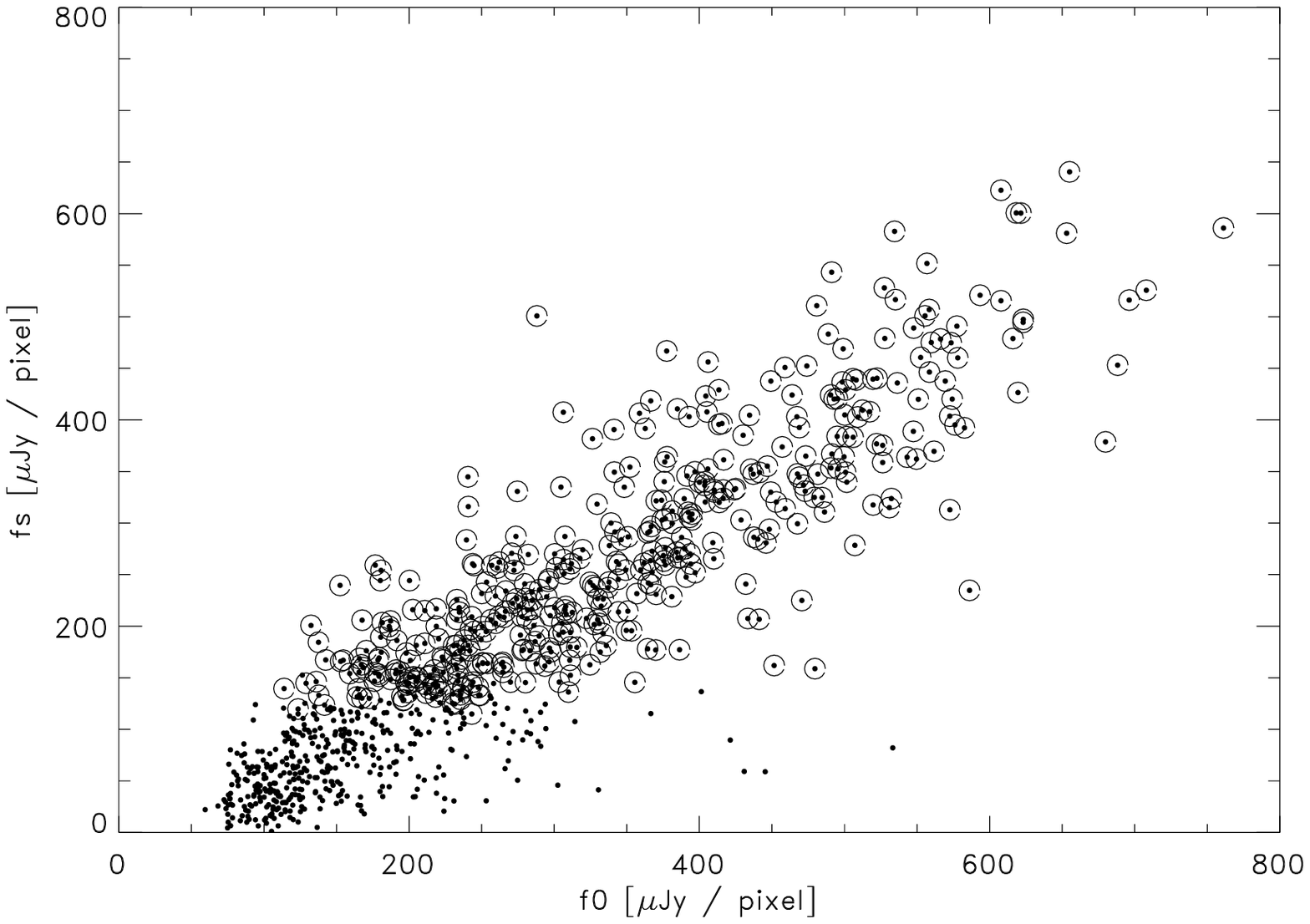,width=9cm}}
\centerline{\psfig{figure=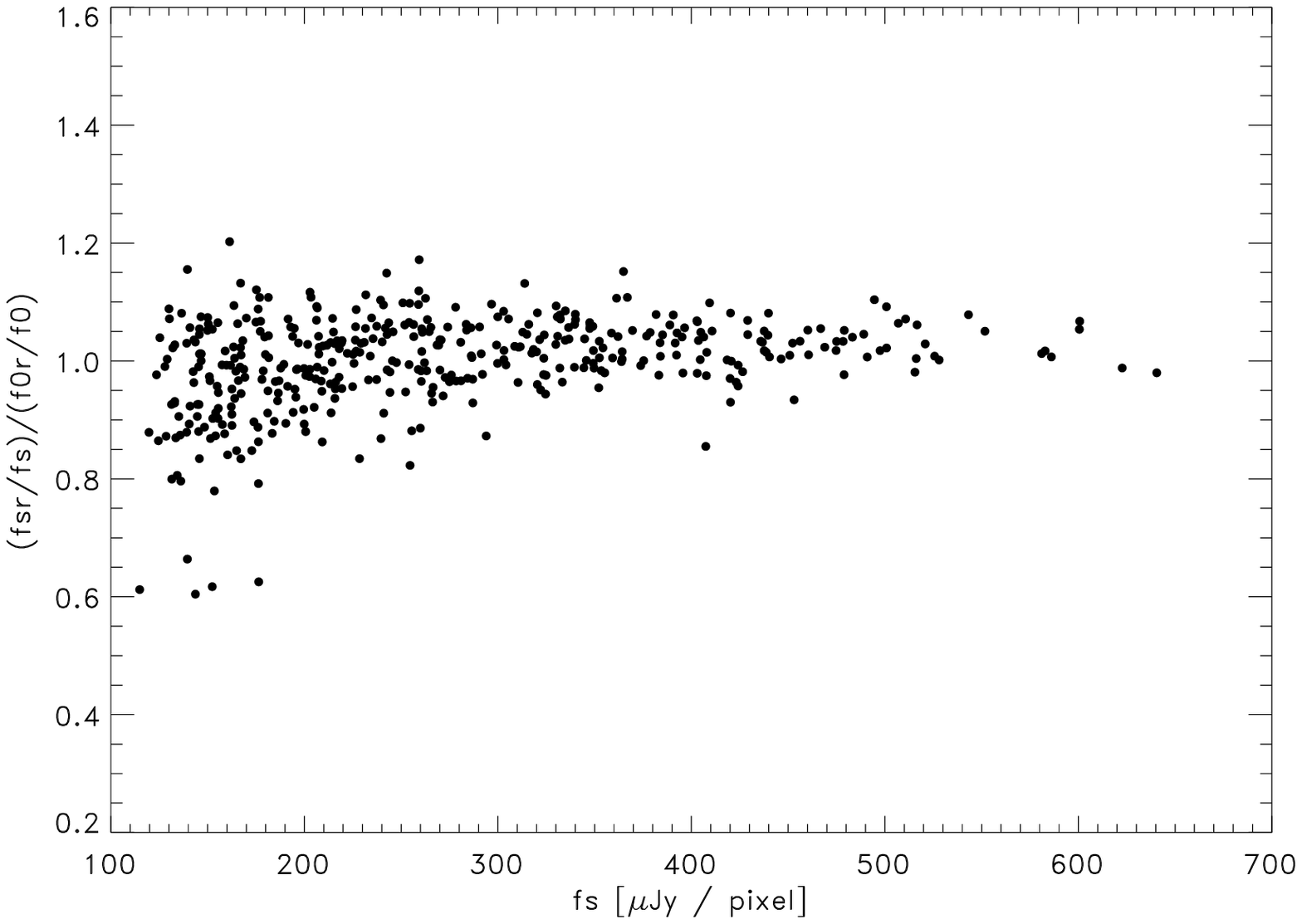,width=9cm}}
\caption{{\it top} - Output flux after reduction versus output flux due only to sampling effects for 
simulated sources. The fluxes are {\bf `unreconstructed'} for transients. The open circles
represent the sources detected above 5 $\sigma$, while the dots are the sources below the 5 $\sigma$
threshold. {\it bottom} - {\bf `Reconstructed'}-to-{\bf `unreconstructed'} flux ratio ($fsr/fs$) normalized to the same 
ratio obtained for mapping effects only ($f0r/f0$) as function of the detected {\bf `unreconstructed'}
peak flux. The ratio distribution broadens towards lower values at faint fluxes, due to the
characteristic of {\it LARI method} of not reconstructing faint sources.}  
\label{fs_f0}
\end{figure}

By comparing the output fluxes (per pixel) obtained for the simulated sources affected only by the 
mapping effects ($f0$) with the output fluxes of the reduced simulated sources ($fs$), we find a  
correlation (although not a 1 to 1 correlation, since the reduced fluxes are always slightly lower 
than the unreduced ones, see figure \ref{fs_f0}({\it top})). A similar correlation is observed for the 
corresponding
{\bf `reconstructed'} peak fluxes ($f0r$ and $fsr$, reconstructed from the transients), although
for faint sources our algorithm is not able to reconstruct correctly the fluxes (see figure
\ref{fs_f0}($bottom$)). 

\nin The dispersion of values in figure \ref{fs_f0}({\it top}) is caused by two kind of errors:
\begin{itemize}
\item[1)] an error proportional to the flux caused by the reduction method limits
or by the mapping and finite integration time effects. In either cases, 
this error affects the peak fluxes.
\item[2)] an additive error caused by the presence of noise and confusion. This error is more
effective at low fluxes than at high ones.
\end{itemize}
At low fluxes, the combination of these errors may cause the total loss of a source 
(i.e. incompleteness).

\begin{figure}
\centerline{\psfig{figure=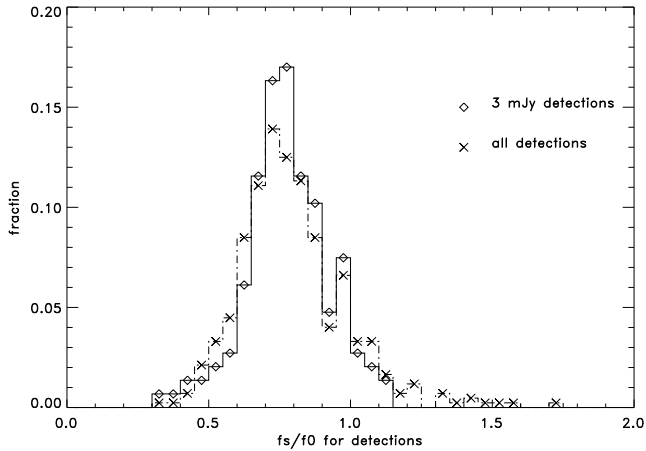,width=9cm}}
\caption{Distribution of the ratio between the reduced peak flux and
the unreduced peak flux (derived from mapping effects only), $fs/f0$, for all the 424 simulated 
sources detected above $5 \sigma$ (dot-dashed line marked by diagonal crosses) and for the 147 
detections injected at 3 mJy (solid line marked by diamonds).}
\label{fs_over_f0}
\end{figure}

\begin{figure}
\centerline{\psfig{figure=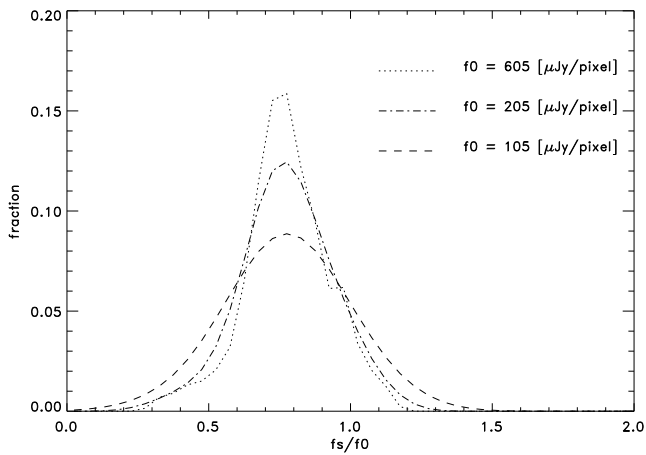,width=9cm}}
\caption{Predicted distributions of the ratio between the reduced peak flux and
the unreduced peak flux (from mapping effects only), $fs/f0$, computed
in presence of a noise of 26 $\mu$Jy. These distributions are shown for three different average
values of $f0$ corresponding to different input fluxes (i.e. $<f0> = 605~ \mu$Jy for an input flux 
of 3 mJy).}
\label{fsf0}
\end{figure}

Figure \ref{fs_over_f0} shows the distribution of the ratio between $fs$ and $f0$, for the 424 
simulated sources detected above $5 \sigma$,
compared to the same ratio for the 147 detections at 3 mJy only. As already noted in figure
\ref{fs_f0}($top$), the ratio is not centered on the value of 1, thus causing a general underestimation
of the total fluxes (derived from $fs$). There are two reasons for this effect: one is the fact
that $f0$ is computed on the measured positions of $fs$ (about 11\% higher than on real positions); 
the other is the under-evaluation of the wings of faint sources by the {\it LARI method}.
The former effect is caused by the fact that projection effects cause $f0$ to be 
enhanced at favorable positions (i.e. center of a pixel on the single images), 
affecting also the centroid position even in the absence of noise. This 
overestimates the peaks of simulated sources which do not fall
on the center of a pixel, thus leading to a bias in the ratio between the 
real $f0$ peak flux and the measured $f0$ simulated peak flux. 
The values of $f0$ computed at the positions measured for $fs$ are on average 
$\sim$11\% higher than the ones computed at real positions.\\
The total distribution is larger and with a longer tail than the 3 mJy one. This is caused by
the noise contribution to the errors, which is more significant at low than at high fluxes,
as shown in figure \ref{fsf0}. In this plot, the predicted distributions of the ratio $fs/f0$ 
in presence of a noise of 26 $\mu$Jy are shown for three different values of $f0$, corresponding to
different mean values of $f0$ for different input fluxes (i.e. $<f0> = 605~ \mu$Jy for 3 mJy input 
flux). As we can notice, the presence of noise broadens the flux distributions and this effect
becomes stronger towards fainter fluxes. The predicted and unbiased distribution of $fs/f0$ for
the brighter sources (3 mJy) peaks at 0.78 $\pm$ 0.03, value subsequently assumed to correct our measured 
fluxes (see section \ref{flux}). 

\begin{figure}
\centerline{\psfig{figure=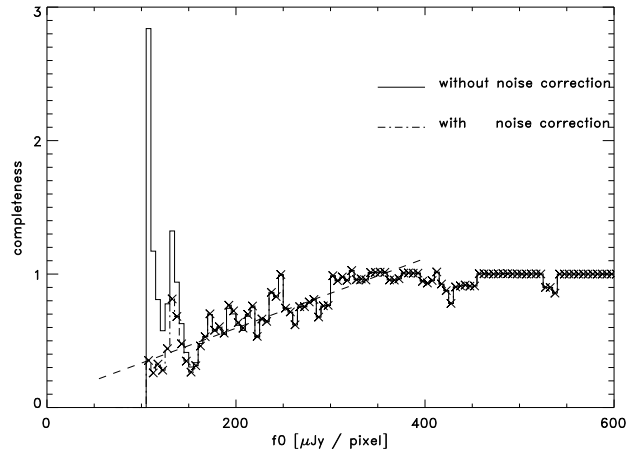,width=9cm}}
\caption{Found-to-predicted detections ratio as a function of $f0$. The predictions have been obtained 
by considering the multiplicative errors only (due to the reduction method 
and mapping, see section \ref{flux}; solid line); the multiplicative errors plus the 
additive noise contribution, assuming a typical noise level on maps of 26 $\mu$Jy (dashed line).}
\label{comp_f0}
\end{figure}

If we assume that the 3 mJy distribution reflects all the multiplicative error components due to the
data reduction and we correct for the noise effect, we can predict the distribution of the 
detection rate for all the total and peak fluxes considered and derived in the simulation.  
Figure \ref{comp_f0} shows the ratio of found-to-predicted detections as a function of $f0$, with the 
predictions obtained by
considering two sources of error: (1) the multiplicative errors only (due to the reduction method and 
mapping, see section \ref{flux}; solid line); (2) the multiplicative errors plus the additive noise 
contribution, assuming a typical noise level on maps of 26 $\mu$Jy (dashed line). 
It is visible a decrease of detection rate/predicted rate below $\sim$300 $\mu$Jy/pixel, which 
corresponds roughly to 1.5 mJy in total input flux. This deficiency is the nominal incompleteness
of our data reduction method.

\begin{figure}
\centerline{\psfig{figure=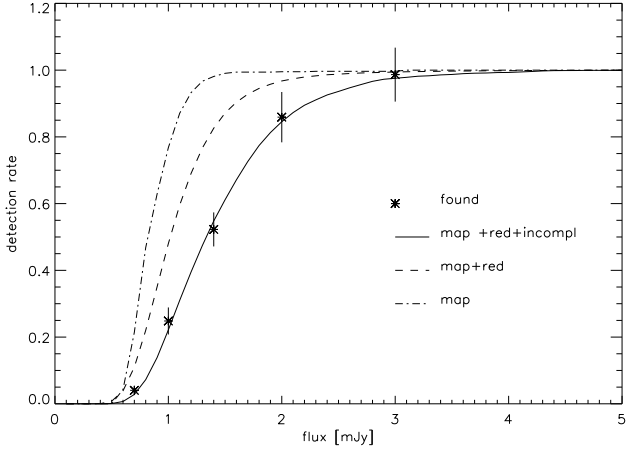,width=9cm}}
\caption{Percentage of detected sources in simulations of ELAIS S1 as a function of the source
input flux. As detection threshold we have considered 5$\sigma$. The dash-dotted curve 
represents the effect of PSF sampling on the detection rate; the dashed curve 
represents the effect of the {\it LARI method}; the solid curve is the total effect on
reduced LW3 ELAIS data.}
\label{fig_compl}
\end{figure}

The detection rates at different fluxes derived with our simulations are reported 
in Table \ref{tab_compl}, where the predictions if only mapping smearing would be present, the
predictions considering the data reduction but without taking into account the incompleteness of our 
method (see figure \ref{comp_f0}), the predictions considering also the incompleteness curve and the
found values are given as detection fractions (i.e. detected sources / input sources) for the five 
different input fluxes.
In figure \ref{fig_compl} the detection rate curves relative to the values given in Table 
\ref{tab_compl} are shown as function of the input flux.

As shown both in Table \ref{tab_compl} and in figure \ref{fig_compl}, almost all the injected 3 mJy sources are detected
at $\geq$ 5$\sigma$. Thus, the detection rate is 98.7\% above 3 mJy, and it remains above 85.9\%
at 2 mJy. However, the detection rate drops quickly at fainter fluxes, in fact it reaches 52.3\%
at 1.4 mJy and 24.8\% and 4.4\% at 1 and 0.7 mJy respectively. Both sampling and reduction method 
are responsible for the large source undetectability at faint fluxes,
although the contribution due to {\it LARI method} seems to become more important around 1.4 - 
1 mJy, then it keeps almost constant, while the PSF sampling effect significantly 
decreases the detection rate for fluxes fainter than 1 mJy.

\begin{table*}
\begin{minipage}{170mm}
\centering
  \caption{Detection rates}
  \label{tab_compl}
\begin{tabular}{ccccccccc}
 &  \\ \hline
input flux & \multicolumn{2}{c}{predicted(mapping)} & \multicolumn{2}{c}{predicted(map+reduction)} &
 \multicolumn{2}{c}{predicted(map+red+incompl)} & \multicolumn{2}{c}{detected} \\
  (mJy)    & over injected & (\%)  & over injected & (\%) & over injected &  (\%) & over injected & (\%)   \\  
 \hline
 0.7  &    42.5/198 &   21.5  &   21.8/198 &  11.0 &    5.5/198 &    2.8  &     8/198 &   4.0 \\ 
 1.0  &   114.4/149 &   76.8  &   71.5/149 &  48.0 &   33.0/149 &   22.1  &    37/149 &  24.8 \\
 1.4  &   195.2/199 &   98.1  &  164.4/199 &  82.6 &  109.0/199 &   54.8  &   104/199 &  52.3 \\
 2.0  &   148.3/149 &   99.5  &  144.2/149 &  96.8 &  126.0/149 &   84.6  &   128/149 &  85.9 \\
 3.0  &   148.5/149 &   99.6  &  148.2/149 &  99.4 &  145.4/149 &   97.6  &   147/149 &  98.7 \\
\hline
\end{tabular}
\end{minipage}
\end{table*}

It must be pointed out that these incompleteness factors cannot be directly translated to the 
real catalogue, which has not a monochromatic flux distribution as the simulations. These factors can
however be used to obtain the completeness of the catalogue and the source counts corrections,
assuming a model for the $log N - log S$ (see Gruppioni, Lari, Pozzi et al. 2001).
When applying simulations to real sky maps we must also remember that there are other sources of error
not included in the simulations, as flat field corrections and distortion tables. The latter
causes a higher smearing of the images and a larger uncertainty on centroid positions.

\section{Flux determination}
\label{flux}

The simulation procedure described in section \ref{autosimulazione} has been used also to 
estimate the effective PSF on a source, real or simulated, and its total 
flux, given its position and peak flux. This procedure, performed on sources to determine their
total flux from their peak flux and position is called ``auto-simulation''.

The linear relation existing between $fs$ and $f0$ (see figure \ref{fs_f0}($top$)) allows
the definition of a flux estimate for both simulated and real sources:
\begin{equation}
f_{total} = fs \times (flux / f0) \label{f_total}. 
\end{equation}
While for simulations $flux$ is the injected total flux, 
for real sources we need to adopt a rough estimate for $flux$ to derive $f_{total}$. $Flux$ 
is the total injected flux used to compute $f0$ (for simulations is 0.7, 1, 1.4, 2 and 3 mJy).
Since transient corrections depend (slowly) on $flux$, for real sources we started with a
rough estimate for $flux$ and then we iterated equation \ref{f_total} to obtain a good
estimate of $f_{total}$ also for strong sources. \\
The starting rough estimate of $flux$ is obtained by:
\begin{equation}
flux = fs / < fs / flux >_{sim} \label{flux}
\end{equation}
where $< fs / flux >_{sim}$ = 0.132 was the average value taken from simulations.\\
Given this input total flux, we can derive $f0$ for real sources exactly as we did for 
simulated sources (see section \ref{autosimulazione}).  
Then, by using formula \ref{f_total}, corrected for the systematic bias of the $fs/f0$
distribution (i.e. divided by 0.78; see section \ref{simulazione}), we obtain the value of
the total flux, $f_{total}$, for our sources. 
Given the relation \ref{f_total}, figure \ref{fs_f0} could also be seen as the representation
of the $f_{total} / flux$ (i.e. measured flux / true flux) distribution.

\begin{figure}
\centerline{\psfig{figure=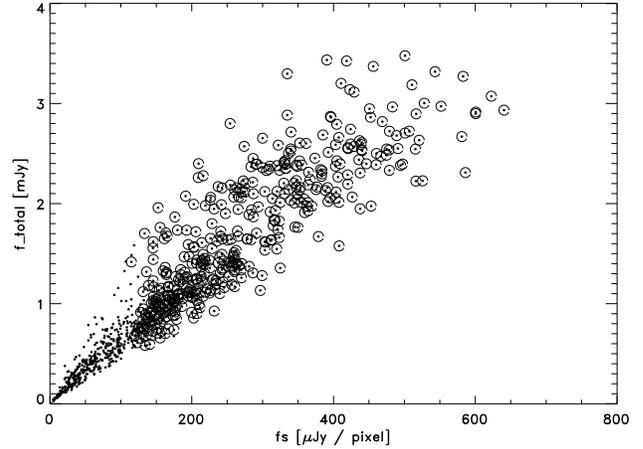,width=9cm}}
\caption{Total fluxes obtained with the auto-simulations versus reduced peak fluxes for simulated sources.
The open circles represent the sources detected above 5 $\sigma$, while the dots are the sources below the 
5 $\sigma$ threshold.}
\label{flussi_fs}
\end{figure}

In figure \ref{flussi_fs} the total fluxes obtained with this procedure are plotted as function of the
reduced peak fluxes obtained for all the simulated sources. As in figure \ref{fs_f0}, the open circles
represent the sources detected above 5 $\sigma$, while the dots are the sources below the 5 $\sigma$
threshold.

The total fluxes obtained with the auto-simulations for the simulated sources are then compared 
with the total fluxes obtained with aperture photometry of the same simulated sources.

\begin{figure}
\centerline{\psfig{figure=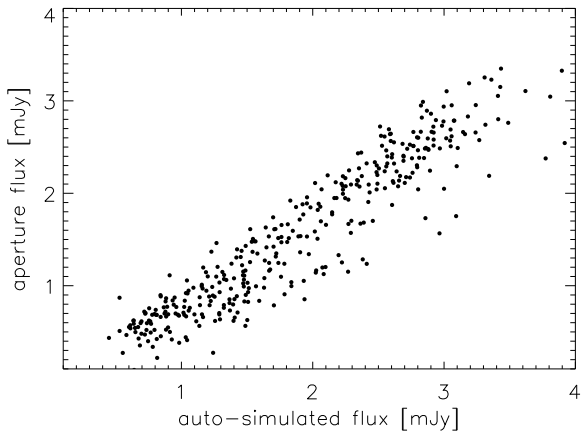,width=9cm}}
\caption{Aperture photometry flux normalised to its relative effective PSF  
versus total flux obtained with auto-simulations for simulated sources.}
\label{fig_aper}
\end{figure}

Concerning the aperture photometry fluxes, we found that the better determination was achieved with 
an aperture radius
of 8$^{``}$, after correcting for the missing flux outside the aperture (40\% in the PSF wings at 
distance $> 8^{``}$).   
We have found a good agreement between the two flux determinations (see figure \ref{fig_aper} for
a comparison between the two results). As total flux estimates for our real data sources, we have 
then decided to adopt the fluxes obtained from the auto-simulations.

In addition to the systematic bias affecting the $f0$ measured values, there is also a 
flux-dependent bias at low signal to noise levels, which derives from the fact that
only sources with a high $fs/f0$ value and with positive noise fluctuations can 
be detected. However, only constant bias corrections are applied to our catalogue data. 

\subsection{Flux errors}

As already mentioned in section \ref{simulazione}, there are two main sources of uncertainty on our source
fluxes: a multiplicative error due to mapping and data reduction method and an additive error
due to the presence of noise in the map (neglecting the uncertainties due to flat-fielding corrections and
field distortions, the latter always depressing the peak fluxes).
As mentioned in section \ref{autosimulazione}, the spread in the $f0$ distribution caused by position
errors is about 14\%. This spread is not only an important cause of the total flux bias, but it contributes
significantly to the width of the $fs/f0$ distribution ($\simeq$0.18) at high fluxes. The extra
contribution from data reduction is about 11\%.
Since simulations show that this spread is rather insensitive of fluxes, we assumed that the 
multiplicative error is constant.

Because our total fluxes are obtained from the ratio between peak fluxes and auto-simulated peak fluxes, 
the combination of the two errors leads to a flux-dependent distribution like 
the one shown in figure \ref{comp_f0}.

Being the width of the $fs/f0$ distribution equal to 0.18 at high signal to noise 
levels, the distribution convolved with noise will have a width:

\begin{equation}
w^2=(0.18)^2+(rms/f0)^2
\end{equation}

We used this relation to obtain the relative flux errors for sources.

\section{Test of the photometry}
\label{calibration}

The photometric accuracy of our reduction of the S\_1 area can be
tested using the stars of the field. Aussel \& Alexander (in prep.;
see also Alexander \& Aussel, 2000) have
performed a detailed study of the mid-infrared emission of stars, from
large sample of sources drawn from the IRAS Faint Source Calalog with
excellent counterparts in the Tycho-2 catalog (Hog et al., 2000). They
show that the B-V color of stars is extremely well correlated with the
B-[12] color, where [12] is a magnitude scale constructed from the
IRAS flux, following the prescriptions of Omont et al. (1999). This
relation allows to predict accurately the 12 $\mu$m IRAS flux of a star,
provided that its B-V is known, and is lower than 1.3. Stellar
atmosphere models (Lejeune et al., 1998) show that for the spectral
types hotter than K3 that the color criteria select, the ratio of the
15 $\mu$m flux to the 12 $\mu$m flux of stars is constant. We have
therefore used the relation calibrated on IRAS data by Aussel \&
Alexander to predict the fluxes of stars in the field, and we compare
them to the product of our analysis. 

\begin{figure}
\centerline{\psfig{figure=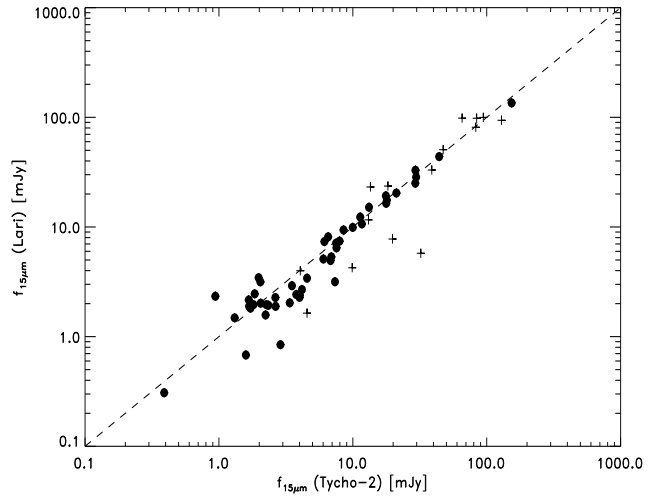,width=8.5cm,height=7cm}}
\caption{LW3 15 $\mu$m fluxes measured with our analysis for the 63 stars 
from the Tycho-2 catalogue in the S1 area versus 15 $\mu$m fluxes
predicted using the relation calibrated on IRAS data by Aussel \&
Alexander. Filled circles are the 48 stars with $B-V < 1.3$, while
crosses are the 15 stars with $B-V > 1.3$. The dashed line 
shows the one-to-one relation.}   
\label{fig_tycho}
\end{figure}

\begin{figure}
\centerline{\psfig{figure=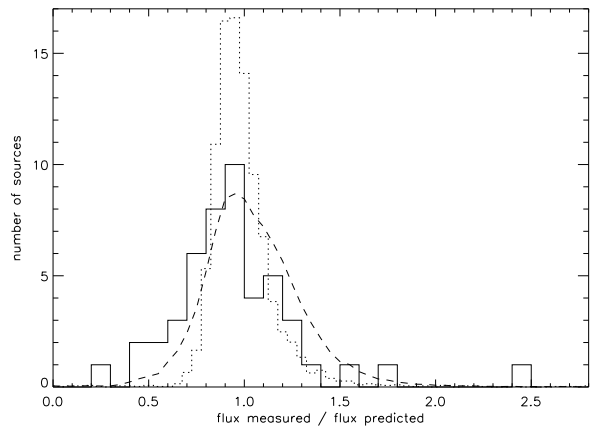,width=8.5cm}}
\caption{Measured to predicted flux distribution for a sample of 3950 stars 
from the study of Aussel \& Alexander (IRAS 12 $\mu$m fluxes, dotted line),
predicted distribution on S\_1 (dashed line) and the distribution of the 
48 stars from the Tycho-2 catalogue detected in our analysis of S1 (LW3 15 $\mu$m 
fluxes).}
\label{fig_calib}
\end{figure}

The area surveyed in S\_1 contains 170 stars in the Tycho-2 catalogue,  
145 of which with $B-V < 1.3$.
In our analysis we detect 63 of them, 48 with $B-V < 1.3$. 
We plot on figures \ref{fig_tycho}
and \ref{fig_calib} respectively the measured fluxes versus
the predicted fluxes and  
the histogram of the ratio of the measured flux to the predicted flux. 
In figure \ref{fig_tycho} the dashed line shows the one-to-one relation,
followed by our data over more than two order of magnitude in flux.
In figure \ref{fig_calib}, the dotted line shows the ratio of measured 
over predicted IRAS 12 $\mu$m
fluxes, for a sample of 3950 stars from the study of Aussel \&
Alexander. The distribution is a skewed log-normal, dominated by the error 
of the IRAS FSC photometric error of the order of 10\% on average. It is
skewed toward observed fluxes higher than predicted fluxes, because
some stars present an excess of IR emission due to the presence of a
disk or shells. Dashed line is the result of the covolution of the former 
distribution with the $fs/f0$ strong sources distribution to simulate 
the spread of values we would expect in our analysis, neglecting noise. 
The mean value of this distribution is 1.047 while the 48 stars in S\_1
with $B-V < 1.3$ have a mean value of .955 leading to a relative flux
scale of  $1.096\pm.044$.
 
The solid line is the ratio of the measured LW3 flux 
and predicted fluxes for the 48 stars detected in S\_1. 

The shape of
the distribution is the same as the dashed one, apart the small scale factor,
with the same skewness and we are confident our fluxes are correct, over a 
large range of fluxes since these stars go from 0.85 mJy to 135 mJy in LW3.

\section{Positional accuracy}
\label{astrometry}

The positional errors in RA and DEC for our sources can be considered as the combination of three
different sources of uncertainty: the finite spatial sampling ($\sigma_s$), the reduction method 
($\sigma_r$), and the uncertainties in the pointing accuracy ($\sigma_p$). The latter is due to 
errors in the ISOCAM lens position (the wheel jitter) and results in an offset of about 1.2 pixels 
from the optical axis, that translates to $\sim$7 arcsec with 
a pixel size of 6$^{\prime \prime}$. Moreover, ISO absolute pointing accuracy is about 3 arcsec. 
 
\begin{figure}
  \centerline{\psfig{figure=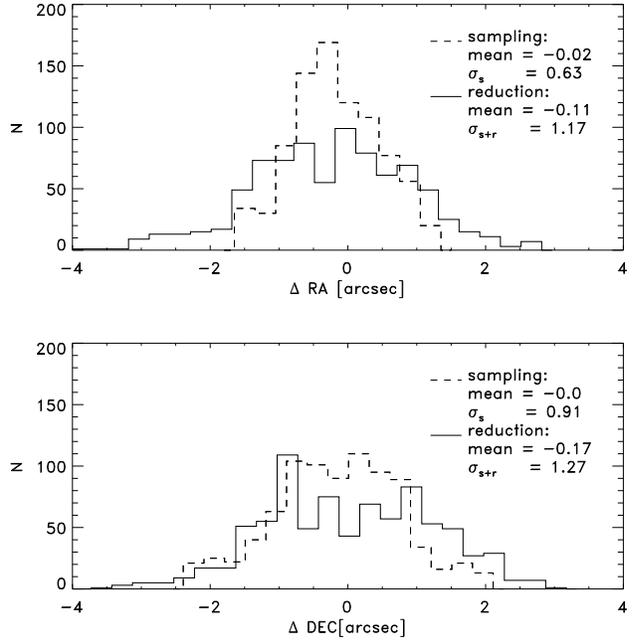,width=9cm} }
  \caption[contours]{Distribution of the difference in RA ({\it top}) and DEC ({\it bottom}) between 
the injected and the found positions for the simulated sources. Dashed line: sampling effect; solid line: sampling plus reduction effect.}
\label{raistofig}
\end{figure}

The effect of the finite spatial sampling ($\sigma_s$) has  been evaluated from the ``theoretical'' simulation 
(see \ref{autosimulazione}), considering the distribution of the differences between the positions of 
the injected 
sources (RA,DEC) and the positions of the (same) sources detected in the projected map (RA0, DEC0).\\
The sum of this effect plus that produced by the method of reduction ($\sigma_{s+r}$) has been evaluated 
from the ``real'' simulation (see \ref{simulazione}), considering the distribution of the differences 
between the positions of the injected sources (RA,DEC) and the positions of the sources detected in the 
projected map after the reduction (RAS, DECS). The widths of these distributions are 0.63 (RA) and 
0.91 arcsec (DEC) for sampling only and 1.17 (RA) and 1.27 arcsec (DEC) for sampling and reduction
effects.
In figures \ref{raistofig} we plot the distributions of the differences in RA and
DEC between the injected and detected positions. 

\begin{figure}
  \centerline{ \psfig{figure=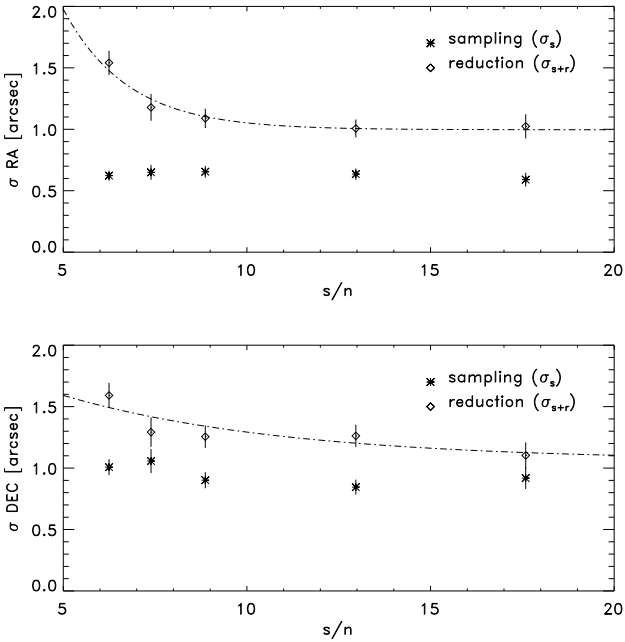,width=9cm} }
 \caption[contours]{Positional errors versus signal-to-noise ratio ({\it top}: RA; {\it bottom}: DEC).
Asterisks represent the errors due to mapping effects only, while diamonds are the errors
due to mapping and data reduction effects.
The dot-dashed lines represent exponential fits to the error dependency on source signal-to-noise ratio.}
\label{rmsradecfig}
\end{figure}

By using our simulations at different input fluxes, we have also checked the dependence of the 
positional errors on source signal-to-noise ratio, as shown in figure \ref{rmsradecfig}. While the 
positional accuracy due to sampling only is almost constant with signal-to-noise 
($\sigma_{s}\approx$0.9$^{\prime\prime}$ for DEC and $\sigma_{s}\approx$0.65$^{\prime\prime}$ for RA), 
as expected being a pure geometrical factor, the positional accuracy 
after the reduction is strongly dependent on signal-to-noise, increasing by about
50\% from $S/N \geq 10$ to $S/N \approx 5 - 7$ 
(i.e. $\sigma\approx$1.0$^{\prime\prime}$ for RA at $S/N \geq 10$; $\sigma\approx$1.5$^{\prime\prime}$ at $5 \leq S/N \leq 7$).
These dependences can be approximated by exponential laws of the form :
\begin{equation}
\sigma_{s+r}(RA) =1.0 + 17.17 \times e^{-(0.57 \times S/N)}
\end{equation}
\begin{equation}
\sigma_{s+r}(DEC) = 1.06 + 1.21 \times e^{-(0.16 \times S/N)}
\end{equation} 
\nin 
These laws, plotted in figure \ref{rmsradecfig} as solid lines and found with a non-linear least squares fit,  
have been used to estimate the positional errors due to the mapping and reduction method as a function, 
for each source, of its signal-to-noise.

The errors introduced by uncertainties in the ISOCAM pointing can be estimated by
performing optical identifications for the sources found in each raster 
and computing an offset with respect to the optical astrometric reference system. 
As optical reference list we have used the PMM USNO-A2.0 Catalogue (Monet 1998).

With between 28 and 36 ISOCAM/USNO coincidences per raster found, we derived the 
median offsets for each of the 11 frames using the following procedure.
First, we have cross-correlated the ISOCAM and USNO lists using a maximum distance of 60 arcsec 
to obtain the best value for the maximum distance for reliable identification.
This distance resulted to be 12 arcsec for all the rasters.
Then, for each raster we have obtained the median offset values in RA and DEC ($med_{ra1}$ and 
$med_{dec1}$) using all the ISO-USNO sources with a maximum distance of 12 arcsec.
We applied the offset values to the ISO positions and we have cross-correlated again the ISO and the
USNO catalogues.

\begin{table}
\centering
  \caption{ISO-USNO astrometric corrections}
  \label{offstab}

\begin{tabular}{lccc}
 &  \\ \hline
Raster & Nominal Position & RA ($^{\prime \prime}$) & DEC ($^{\prime \prime}$) \\
       &  (J2000) & offset ~~~error & offset ~~~error \\  
 \hline
S1$\_$1  & 00 30 25.4 $-$42 57 00.3 & $-$2.06 $\pm$ 0.40 & $-$4.46 $\pm$ 0.38 \\ \hline
S1$\_$2  & 00 31 08.2 $-$43 36 14.1 & $-$3.24 $\pm$ 0.22 & +6.86 $\pm$ 0.29 \\ \hline
S1$\_$3  & 00 31 51.9 $-$44 15 27.0   & +1.57 $\pm$ 0.29 & $-$7.75 $\pm$ 0.33 \\ \hline
S1$\_$4  & 00 33 59.4 $-$42 49 03.1   & +0.23 $\pm$ 0.22 & $-$4.01 $\pm$ 0.27 \\ \hline
S1$\_$5A & 00 34 44.4 $-$43 28 12.0  & $-$3.50 $\pm$ 0.23 & +9.63 $\pm$ 0.27 \\ \hline
S1$\_$5B & 00 34 44.4 $-$43 28 12.0  & $-$0.52 $\pm$ 0.21 & $-$8.10 $\pm$ 0.26 \\ \hline
S1$\_$5C & 00 34 44.4 $-$43 28 12.0  & $-$3.04 $\pm$ 0.24 & +5.34 $\pm$ 0.29 \\ \hline
S1$\_$6  & 00 35 30.4 $-$44 07 19.8   & +0.60 $\pm$ 0.43 & $-$7.14 $\pm$ 0.24 \\ \hline
S1$\_$7  & 00 37 32.5 $-$42 40 41.2   & +1.26 $\pm$ 0.22 & $-$5.62 $\pm$ 0.38 \\ \hline
S1$\_$8  & 00 38 19.6 $-$43 19 44.5   & +0.72 $\pm$ 0.22 & $-$5.31 $\pm$ 0.24 \\ \hline
S1$\_$9  & 00 39 07.8 $-$43 58 46.6 & $-$2.34 $\pm$ 0.19 & +4.61 $\pm$ 0.23 \\ \hline
\end{tabular}
\end{table}

We have selected again all the sources within a maximum distance of 12 arcsec and we have used these 
sources to calculate new median offset values ($med_{ra2}$ and $med_{dec2}$).
The total offsets for each raster have then been obtained as $RA\_offset = med_{ra1} + med_{ra2}$,
$DEC\_offset = med_{dec1} + med_{dec2}$. The values of the offsets and their relative errors (computed
as the standard errors on median: $\sigma_{med}=1.2533\frac{\sigma}{\sqrt{N}}$ (Akin \& Colton
1970), where $\sigma$ and $N$ are 
respectively the standard deviation and the number of sources considered in each raster) have been reported in 
Table \ref{offstab}.
Each source position has then been corrected for the offset found for the corresponding 
raster. The error ($\sigma_p$) introduced on source positions by the presence of the systematic 
offset is given by the error on the offset determination (see column 4 and 6 in Table \ref{offstab}).
This error has been added to the positional uncertainty due to mapping and reduction method
($\sigma_{s+r}$) to obtain the total position error for each source:
\begin{equation}
\sigma^2_{RA} = \sigma^2_{s+r}(RA) + \sigma^2_{p}(RA)  
\end{equation}
\begin{equation}
\sigma^2_{DEC} = \sigma^2_{s+r}(DEC) + \sigma^2_{p}(DEC)
\end{equation}

\begin{figure*}
\begin{minipage}{170mm}
  \centerline{\psfig{figure=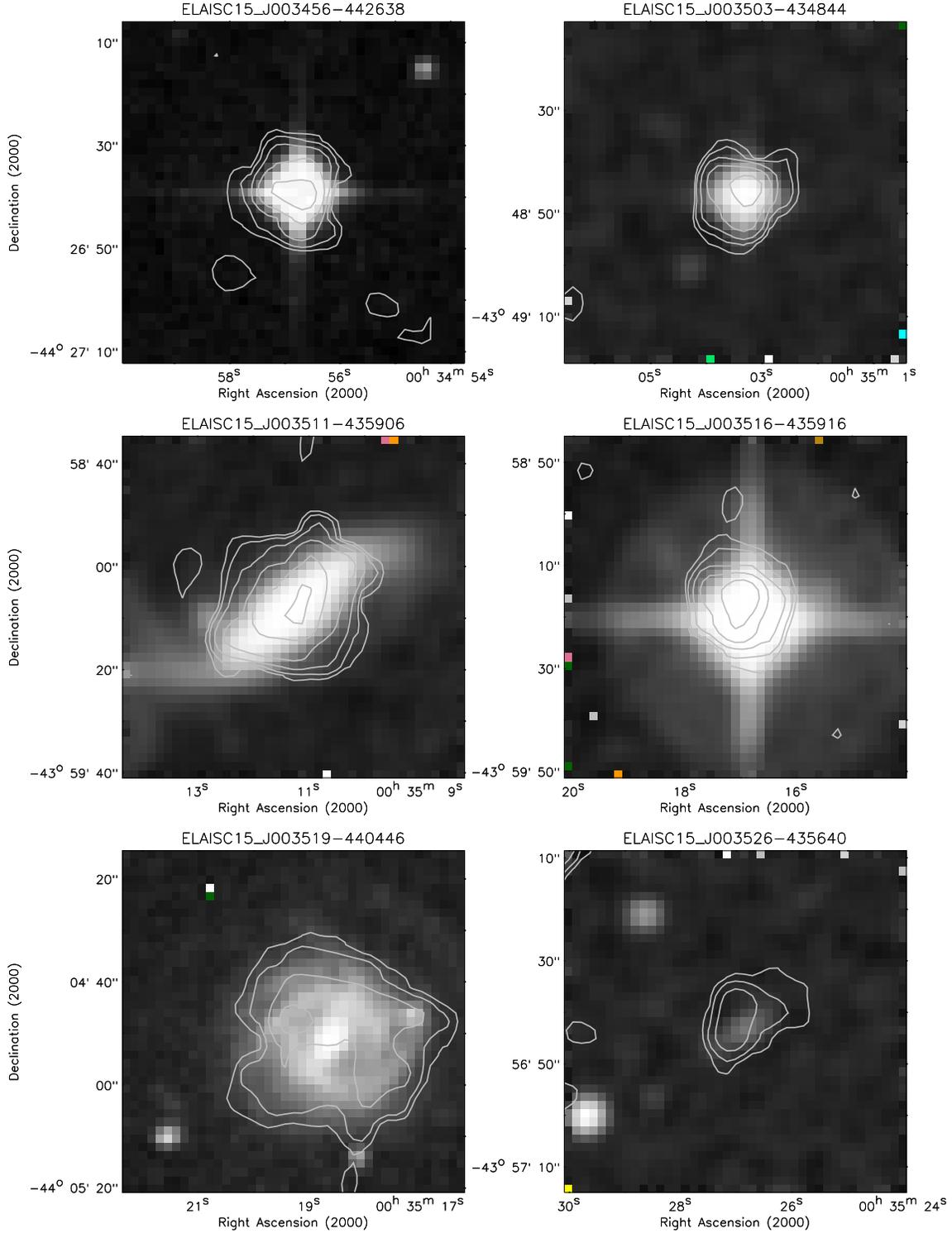,width=20cm} }
 \caption[contours]{Example of ISOCAM 15 $\mu$m contours superimposed to DSS optical images
($b_j$ band). Contour levels of the 15 $\mu$m emission correspond to 0.05, 0.08, 0.15, 0.3, 0.6, 1.2, 2.0,
4.0 mJy / pixel. The size of each image is 1$\times$1 arcmin.}
\label{fig_contour}
\end{minipage}
\end{figure*}

In figure \ref{fig_contour} is shown an example of our ISOCAM 15 $\mu$m contour levels superimposed 
to DSS optical images, after correcting for the systematic offsets computed above. 
The plot can give an idea of the astrometric accuracy of our catalogue and images. In fact, as 
clearly visible from the figure, the positions of our sources after offsetting appear as accurate 
as we estimated from simulations (see above), thus allowing reliable identifications within 
a few arcseconds or less.
 
\section{Repeated central region, S1$\_$5}

The central field of the S1 area, as mentioned in section \ref{mapcreation}, has been observed 
three times in order to reach a deeper flux limit with respect to the rest of the area and to 
allow reliability checks on sources. 

The reduction of the three observations was carried out in the standard way (see section \ref{data_reduction}) 
until the stage of map creation. After the creation of the three single raster maps, some particular 
routines have been applied for combining them and for performing simulations in the combined 
map. \\
To obtain a unique combined map from the three single observations, first we have projected 
all of them on the sky with the same orientation (north-south). The three single rasters have 
then been corrected for the relative astrometric offsets (see section \ref{astrometry}) and
then coadded. For the coaddition, each raster map has been weighted, with a weight 
proportional pixel-by-pixel to its relative ``NPIX'' map. The combined ``NPIX'' map was just 
the pixel-by-pixel sum of each ``NPIX'' map.
The ``RMS'' distribution over the mosaic map has been obtained with the standard procedure 
(see \ref{data_reduction}). The average rms value in the central part of the
combined S1$\_$5 map is about 0.016 mJy.

Once obtained the combined map, source extraction have been performed in the same way 
as for the single observation raster maps (see section \ref{source_ext}). 
In S1$\_$5 we have detected 93 sources.

To derive the total fluxes from the detected peak fluxes, we have performed the
``auto-simulations''
(see section \ref{flux}) for the 93 sources, by injecting point sources into each of the three
single fields and then combining the resulting images with the same weight used to coadd 
the real maps. The total-to-peak ratio found for the simulated sources have then been
applied to the peak flux obtained for the real sources to get their total flux, exactly
in the same way as we did for the single rasters in the rest of the S1 area.

Total fluxes in the combined map range between 0.57 and $\sim$100 mJy. 

\subsection{Simulations in S1$\_$5}  
 
To perform simulations of the repeated raster we have used and appropriately combined the 
simulations performed separately on the three single rasters, S1$\_$5, S1$\_$5$\_$B, 
S1$\_$5$\_$C (section \ref{simulations}). 
The positions of the sources injected in each of the three individual fields have been 
chosen in order to simulate the effects caused by the application, in the coaddition phase, 
of a relative astrometric offset among the rasters. 50 sources for each of the above 5 total
fluxes (0.7, 1, 1.4, 2 and 3 mJy) were injected. The ``auto-simulations'' were
performed on the positions found on the combined map.

\begin{figure}
\centerline{\psfig{figure=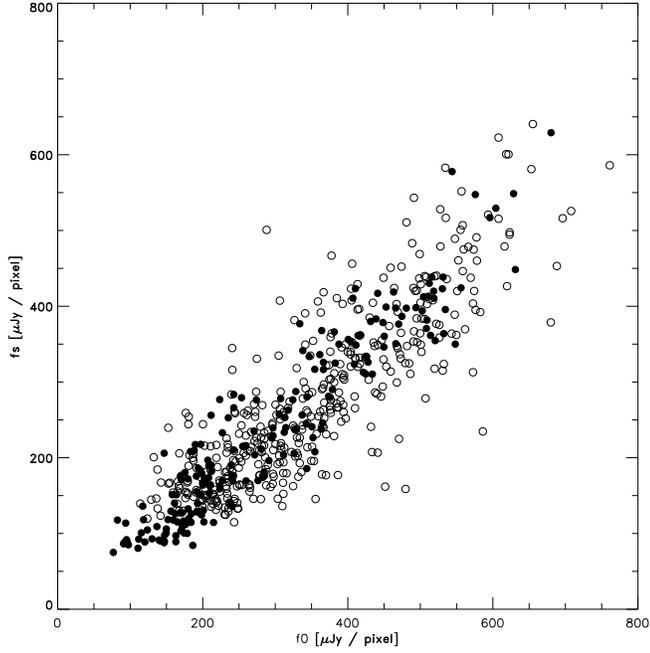,width=9cm}}
\caption{Output flux after reduction versus output flux due only to sampling effects for 
simulated sources detected in the combined map S1$\_5$ (dot) and in the main S1 survey (open circle).}  
\label{fs_f0_mos}
\end{figure}

In figure \ref{fs_f0_mos} the $fs$ and $f0$ peak fluxes of the detected sources are plotted superimposed 
to the ones found for the simulated sources in the main S1 area. Apart from the
deeper detection level, the general trend is the same with a somewhat smaller dispersion.
For the 3 mJy input sources the $fs/f0$ observed distribution peaks at 0.82 $\pm$ 0.03 and its 
width is 0.11, while the corresponding values for single fields simulation are 0.78 $\pm$ 0.03 and 0.18.
  
Following the same procedure as before we can predict completeness and detection rates also for sources in the combined map.

\begin{figure}
\centerline{\psfig{figure=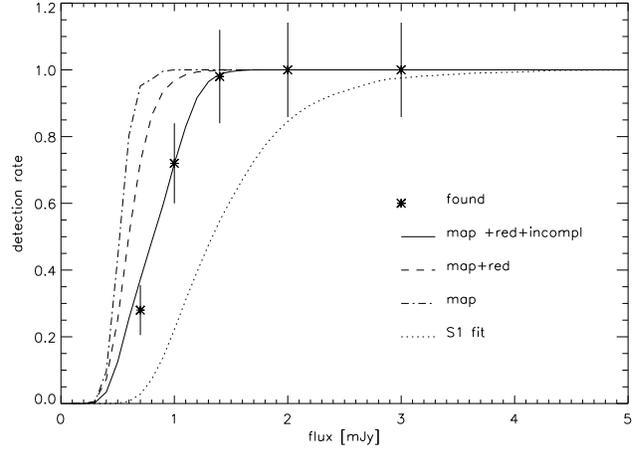,width=9cm}}
\caption{Percentage of detected sources in simulations of ELAIS S1$\_5$ as a 
function of the source
input flux. As detection threshold we have considered 5$\sigma$. The dashed curve 
represents the effect of PSF sampling on the detection rate; the dash-dotted curve 
represents the effect of the {\it LARI method}; the solid curve is the total effect on
reduced LW3 ELAIS data. The dotted curve is the total detection rate found in S1 (the
same as the solid curve plotted in figure \ref{fig_compl})}
\label{fig_compl_mos}
\end{figure}

The results of the simulations have been reported in table \ref{tab_compl_mos} and shown in figure \ref{fig_compl_mos}.

\begin{table*}
\begin{minipage}{170mm}
\centering
  \caption{Detection rates in the S1$\_5$ combined map}
  \label{tab_compl_mos}

\begin{tabular}{ccccccccc}
 &  \\ \hline
input flux & \multicolumn{2}{c}{predicted(mapping)} & \multicolumn{2}{c}{predicted(map+reduction)} &
 \multicolumn{2}{c}{predicted(map+red+incompl)} & \multicolumn{2}{c}{detected} \\
  (mJy)    & over injected & (\%)  & over injected & (\%) &  &  (\%) &  & (\%)   \\  
 \hline
 0.7  &    47.6/50 &   95.2  &   36.2/50 &  72.4 &   18.6/50 &   37.2  &    14/50 &  28.0 \\ 
 1.0  &    50.0/50 &  100.0  &   48.4/50 &  96.8 &   35.9/50 &   71.9  &    36/50 &  72.0 \\
 1.4  &    50.0/50 &  100.0  &   49.9/50 &  99.9 &   49.4/50 &   98.8  &    49/50 &  98.0 \\
 2.0  &    50.0/50 &  100.0  &   50.0/50 & 100.0 &   50.0/50 &  100.0  &    50/50 & 100.0 \\
 3.0  &    50.0/50 &  100.0  &   50.0/50 & 100.0 &   50.0/50 &  100.0  &    50/50 & 100.0 \\
\hline
\end{tabular}
\end{minipage}
\end{table*}

The combination of the three maps does not only reduce the errors in flux determination and increase
the fraction of detected sources at faint fluxes, but provides more precise positions in the sky.

\begin{figure}
  \centerline{\psfig{figure=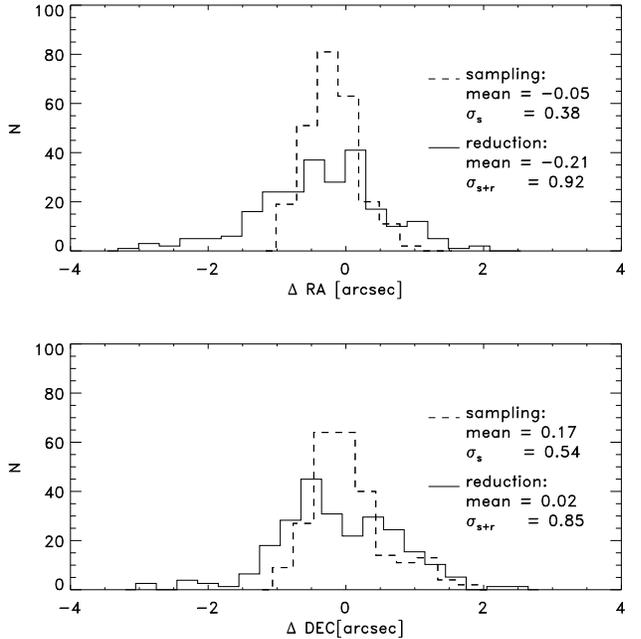,width=9cm} }
  \caption[contours]{Distribution of the difference in RA ({\it top}) and DEC ({\it bottom}) between the injected and the found positions for simulated sources
detected in the S1$\_5$ combined map; dashed line: sampling effect; solid line: sampling plus reduction effect.}
\label{raistomosfig}
\end{figure}

\begin{figure}
  \centerline{ \psfig{figure=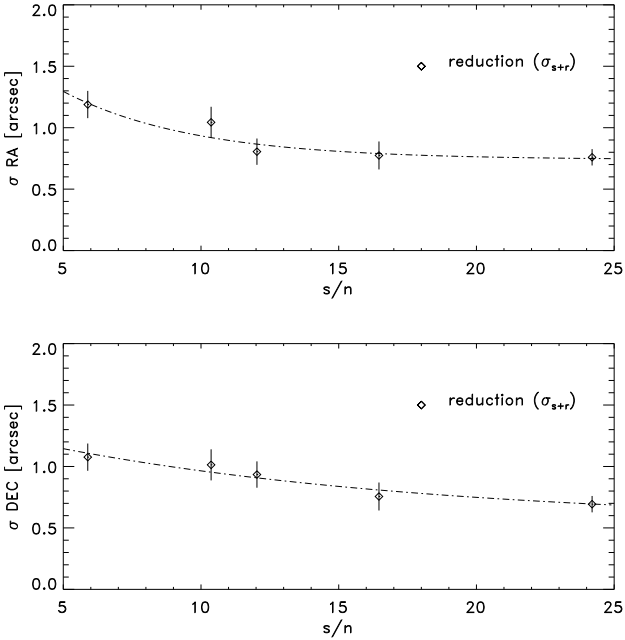,width=9cm} }
 \caption[contours]{Positional errors versus signal-to-noise ratio ({\it top}: RA; {\it bottom}: DEC) for simulated sources detected in the S1$\_5$ combined map. Only
errors due to the combined effect of due to mapping and data reduction are plotted as diamonds.
The dot-dashed lines represent exponential fits to the error dependency on source signal-to-noise ratio.}
\label{rmsradecmosfig}
\end{figure}

Figure \ref{raistomosfig} shows the distributions of the differences in RA and DEC between the injected and detected positions, while figure \ref{rmsradecmosfig}
shows the dependence of position errors on the signal-to-noise ratio. 
The widths found for the distributions in RA and DEC for sampling and reduction effect, are 0.92 and 0.85 respectively, smaller than those found for the main survey, 1.17 and 1.27. Considering the  dependence of the  position errors on the signal-to-noise ratio, the laws found are of the form:

\begin{equation}
\sigma_{s+r}(RA) = 0.74 + 1.59 \times e^{-(0.2 \times S/N)}
\end{equation}
\begin{equation}
\sigma_{s+r}(DEC) = 0.54 + 0.86 \times e^{-(0.07 \times S/N)} 
\end{equation} 

These law are less steep than those found in the all survey and, as we can see from the figure \ref{rmsradecmosfig}, the values of the positional errors near the limit of the survey (signal-to-noise = 5) are, for both coordinates, less than 1.5$^{\prime\prime}$. 

The combination of the three maps, changing the repetition factor from 2 to 6
for each single pointing image,
not only reduces the errors, but also the effects due to mapping.

\section{The Source Catalogue in S1}
The final catalogue obtained with our method contains a total of 462 sources detected at 
15 $\mu$m (LW3) in the ELAIS region S1. All the sources detected over the whole $2^{\circ} 
\times 2^{\circ}$ area have signal-to-noise ratio greater than 5. The catalogue
reports the source name, the offset-corrected position (right ascension and declination at
equinox J2000), the positional accuracy on the final images, the source peak flux (in mJy/pixel), 
the detection level (signal-to-noise ratio), the total flux and its error (in mJy), the raster name
and eventually a note, indicating whether the source is identificated with a star, its flux
have been obtained through aperture photometry, etc.
In case of extended of very bright sources, the total flux reported in the table is computed by
aperture photometry instead of by ``auto-simulations'', the latter method providing a correct 
measurement mainly for unresolved sources.
Note that a few sources (belonging to the border part of a raster, overlapping with an adjacent
raster) might appear in two different rasters. In this case, the repeated sources have been 
reported twice in the catalogue and the corresponding additional raster number is quoted in the 
notes.

As described in section \ref{source_ext}, before extracting sources on the final maps, 
we have identified candidate sources inside the pixel histories (flux excesses above the 
background over the single time-scans greater than 0.7 ADU/gain/s) and on the single calibrated 
images (all pixels with flux $> 0.4$ mJy / pixel), providing 100\% reliable lists of sources above
these two thresholds. Although there is not a perfect correspondence between these two
flux thresholds and total fluxes in the final raster maps (due to flat-fielding, distortions,
etc.), we have found that by splitting the S1 catalogue in two (above and below 1 mJy)
we can provide a highly reliable catalogue, where all the sources have been checked before
extraction and a less reliable but deeper catalogue, where most sources could not be checked
with the above criteria.
The majority of sources fainter than 1 mJy, in fact, might have flux excesses in the single 
pixel histories below 0.7 ADU/gain/s and peak fluxes on the single images fainter than 
0.4 mJy / pixel, the limits chosen for visual inspection, below which is almost impossible 
to distinguish a flux excess on the pixel history from local background fluctuations.
This does not apply to S1$\_5$, because sources have been extracted on the combination of
three single observations, that have been separately checked before the coaddition.

In Tables \ref{catalog_gt1} and \ref{catalog_lt1}, the first page (corresponding to the
first raster) of the catalogues in S1, respectively above and below 
1 mJy, are shown as examples. The full ELAIS S1 + S1$\_5$ catalogues 
at 15 $\mu$m obtained with {\it Lari method} will be 
available from {\it http://boas5.bo.astro.it/$\sim$elais/catalogues/}.

\begin{table*}
\begin{minipage}{180mm}
\centering
  \caption{The 15 $\mu$m $S \geq 1$ mJy Catalogue in the ELAIS Southern Area S1}
  \label{catalog_gt1}
\begin{tabular}{lccccrrrrcl} 
&&&& \\ \hline
\multicolumn{1}{c}{Name} & \multicolumn{1}{c}{RA} & \multicolumn{1}{c}{DEC} & \multicolumn{1}{c}{$\sigma$(RA)} & 
\multicolumn{1}{c}{$\sigma$(DEC)} & \multicolumn{1}{c}{Fpeak} & \multicolumn{1}{c}{$S/N$} & \multicolumn{1}{c}{Ftot} &
 \multicolumn{1}{c}{$\sigma$(Ftot)} & \multicolumn{1}{c}{raster} & \multicolumn{1}{c}{notes}\\
     & \multicolumn{1}{c}{(J2000)} & \multicolumn{1}{c}{(J2000)} & \multicolumn{1}{c}{($\prime\prime$)} &
 \multicolumn{1}{c}{($\prime\prime$)} & \multicolumn{1}{c}{(mJy)} & & \multicolumn{1}{c}{(mJy)} &
 \multicolumn{1}{c}{(mJy)} & & \\
&&&& \\ \hline       
  ELAISC15$\_$J002818$-$424303 &  00 28 18.9 & $-$42 43 03.8 & 1.1 & 1.2  &  0.362  &  14.11 &   2.239  &  0.555 & S1$\_$1 & star      \\
  ELAISC15$\_$J002831$-$425203 &  00 28 31.9 & $-$42 52 03.6 & 1.3 & 1.5  &  0.192  &   7.43 &   1.050  &  0.302 & S1$\_$1 &           \\
  ELAISC15$\_$J002848$-$430658 &  00 28 48.4 & $-$43 06 58.5 & 1.6 & 1.6  &  0.161  &   6.09 &   1.000  &  0.313 & S1$\_$1 &           \\
  ELAISC15$\_$J002853$-$425053 &  00 28 53.9 & $-$42 50 53.7 & 1.7 & 1.6  &  0.158  &   5.62 &   1.041  &  0.355 & S1$\_$1 &           \\
  ELAISC15$\_$J002857$-$425343 &  00 28 57.3 & $-$42 53 43.2 & 1.1 & 1.3  &  0.323  &  10.72 &   2.196  &  0.590 & S1$\_$1 & star      \\  
  ELAISC15$\_$J002904$-$425243 &  00 29 04.4 & $-$42 52 43.1 & 1.5 & 1.5  &  0.175  &   6.62 &   1.087  &  0.322 & S1$\_$1 &           \\
  ELAISC15$\_$J002913$-$431717 &  00 29 13.7 & $-$43 17 17.6 & 1.1 & 1.1  &  0.925  &  37.50 &   7.648  &  1.797 & S1$\_$1 & star      \\
  ELAISC15$\_$J002915$-$430333 &  00 29 15.8 & $-$43 03 33.7 & 1.1 & 1.3  &  0.288  &  12.64 &   1.855  &  0.470 & S1$\_$1 &           \\
  ELAISC15$\_$J002917$-$423921 &  00 29 17.4 & $-$42 39 21.9 & 1.4 & 1.5  &  0.181  &   7.10 &   1.093  &  0.318 & S1$\_$1 &           \\
  ELAISC15$\_$J002930$-$431139 &  00 29 30.9 & $-$43 11 39.9 & 1.6 & 1.6  &  0.157  &   6.02 &   1.220  &  0.379 & S1$\_$1 &           \\
  ELAISC15$\_$J002939$-$430625 &  00 29 39.3 & $-$43 06 25.3 & 1.5 & 1.5  &  0.253  &   6.30 &   3.400  &  0.723 & S1$\_$1 & aper  \\
  ELAISC15$\_$J002943$-$423736 &  00 29 43.7 & $-$42 37 36.8 & 1.8 & 1.6  &  0.198  &   5.47 &   1.963  &  0.651 & S1$\_$1 & star      \\
  ELAISC15$\_$J002949$-$430703 &  00 29 49.1 & $-$43 07 03.0 & 1.3 & 1.5  &  0.211  &   7.60 &   1.191  &  0.349 & S1$\_$1 &           \\
  ELAISC15$\_$J002956$-$424534 &  00 29 57.0 & $-$42 45 34.7 & 1.1 & 1.1  &  0.704  &  25.84 &   4.175  &  0.989 & S1$\_$1 & star      \\
  ELAISC15$\_$J003014$-$430332 &  00 30 14.9 & $-$43 03 32.8 & 1.1 & 1.3  &  0.302  &  11.68 &   2.450  &  0.624 & S1$\_$1 &           \\
  ELAISC15$\_$J003017$-$423721 &  00 30 17.7 & $-$42 37 21.9 & 1.4 & 1.5  &  0.176  &   7.02 &   1.470  &  0.427 & S1$\_$1 & star      \\
  ELAISC15$\_$J003022$-$423657 &  00 30 22.7 & $-$42 36 57.5 & 1.1 & 1.1  &  2.017  &  77.90 &  23.000  &  3.900 & S1$\_$1 & aper  \\
  ELAISC15$\_$J003023$-$424549 &  00 30 23.3 & $-$42 45 49.6 & 1.1 & 1.3  &  0.344  &  12.96 &   2.073  &  0.522 & S1$\_$1 &           \\
  ELAISC15$\_$J003025$-$423855 &  00 30 25.2 & $-$42 38 55.2 & 1.3 & 1.5  &  0.189  &   7.37 &   1.134  &  0.329 & S1$\_$1 &           \\
  ELAISC15$\_$J003039$-$425348 &  00 30 39.6 & $-$42 53 48.3 & 1.1 & 1.3  &  0.341  &  13.35 &   1.980  &  0.497 & S1$\_$1 &           \\
  ELAISC15$\_$J003054$-$430044 &  00 30 54.4 & $-$43 00 44.4 & 1.2 & 1.4  &  0.206  &   8.11 &   1.486  &  0.412 & S1$\_$1 &           \\
  ELAISC15$\_$J003101$-$431733 &  00 31 01.8 & $-$43 17 33.1 & 1.1 & 1.1  &  9.746  & 244.20 & 103.000  & 19.200 & S1$\_$1 & star, aper\\
  ELAISC15$\_$J003104$-$425635 &  00 31 04.8 & $-$42 56 35.1 & 1.1 & 1.3  &  0.289  &  11.01 &   2.382  &  0.620 & S1$\_$1 &           \\
  ELAISC15$\_$J003114$-$424228 &  00 31 14.4 & $-$42 42 28.5 & 1.1 & 1.1  &  0.811  &  30.83 &   5.968  &  1.406 & S1$\_$1 &           \\
  ELAISC15$\_$J003123$-$430939 &  00 31 23.6 & $-$43 09 39.3 & 1.5 & 1.5  &  0.179  &   6.49 &   1.032  &  0.318 & S1$\_$1 &           \\
  ELAISC15$\_$J003133$-$424445 &  00 31 33.5 & $-$42 44 45.7 & 1.1 & 1.2  &  0.571  &  21.42 &   4.318  &  1.034 & S1$\_$1 &           \\
  ELAISC15$\_$J003137$-$425844 &  00 31 37.9 & $-$42 58 44.3 & 1.3 & 1.5  &  0.188  &   7.34 &   1.107  &  0.316 & S1$\_$1 &           \\
  ELAISC15$\_$J003142$-$425642 &  00 31 42.9 & $-$42 56 42.2 & 1.8 & 1.6  &  0.150  &   5.51 &   1.188  &  0.389 & S1$\_$1 &           \\
  ELAISC15$\_$J003151$-$431046 &  00 31 51.0 & $-$43 10 46.6 & 1.4 & 1.5  &  0.163  &   6.64 &   1.167  &  0.344 & S1$\_$1 &           \\
  ELAISC15$\_$J003151$-$424540 &  00 31 51.5 & $-$42 45 40.7 & 1.9 & 1.6  &  0.141  &   5.22 &   1.004  &  0.335 & S1$\_$1 &           \\
  ELAISC15$\_$J003216$-$430432 &  00 32 16.4 & $-$43 04 32.4 & 1.1 & 1.1  &  2.025  &  77.09 &  12.136  &  2.820 & S1$\_$1 &           \\
  ELAISC15$\_$J003223$-$430546 &  00 32 23.9 & $-$43 05 46.1 & 1.3 & 1.5  &  0.197  &   7.90 &   1.259  &  0.352 & S1$\_$1 &           \\
  ELAISC15$\_$J003232$-$431306 &  00 32 32.6 & $-$43 13 06.6 & 2.0 & 1.6  &  0.188  &   5.00 &   1.085  &  0.367 & S1$\_$1 &           \\
  ELAISC15$\_$J003233$-$430632 &  00 32 33.1 & $-$43 06 32.2 & 1.4 & 1.5  &  0.190  &   6.64 &   1.663  &  0.527 & S1$\_$1 &           \\
\hline 
\end{tabular}
\end{minipage}      
\end{table*}

\begin{table*}
\begin{minipage}{180mm}
\centering
  \caption{The 15 $\mu$m $S < 1$ mJy Catalogue in the ELAIS Southern Area S1}
  \label{catalog_lt1}
\begin{tabular}{lccccrrrrcl} 
&&&& \\ \hline
\multicolumn{1}{c}{Name} & \multicolumn{1}{c}{RA} & \multicolumn{1}{c}{DEC} & \multicolumn{1}{c}{$\sigma$(RA)} & 
\multicolumn{1}{c}{$\sigma$(DEC)} & \multicolumn{1}{c}{Fpeak} & \multicolumn{1}{c}{$S/N$} & \multicolumn{1}{c}{Ftot} &
 \multicolumn{1}{c}{$\sigma$(Ftot)} & \multicolumn{1}{c}{raster} & \multicolumn{1}{c}{notes}\\
     & \multicolumn{1}{c}{(J2000)} & \multicolumn{1}{c}{(J2000)} & \multicolumn{1}{c}{($\prime\prime$)} &
 \multicolumn{1}{c}{($\prime\prime$)} & \multicolumn{1}{c}{(mJy)} & & \multicolumn{1}{c}{(mJy)} &
 \multicolumn{1}{c}{(mJy)} & & \\
&&&& \\ \hline       
  ELAISC15$\_$J002929$-$430651 &  00 29 29.8 & $-$43 06 51.5 & 2.0 & 1.6  &  0.111  &   5.13 &   0.666  &  0.238 & S1$\_$1 &           \\
  ELAISC15$\_$J002938$-$424123 &  00 29 38.6 & $-$42 41 23.0 & 1.5 & 1.5  &  0.163  &   6.48 &   0.971  &  0.292 & S1$\_$1 &           \\
  ELAISC15$\_$J003128$-$430747 &  00 31 28.9 & $-$43 07 47.2 & 1.9 & 1.6  &  0.119  &   5.25 &   0.736  &  0.278 & S1$\_$1 &           \\
  ELAISC15$\_$J003144$-$425826 &  00 31 44.9 & $-$42 58 26.8 & 1.9 & 1.6  &  0.136  &   5.21 &   0.768  &  0.257 & S1$\_$1 &           \\
  ELAISC15$\_$J003147$-$423548 &  00 31 47.3 & $-$42 35 48.8 & 1.6 & 1.6  &  0.154  &   6.01 &   0.836  &  0.262 & S1$\_$1 &           \\
  ELAISC15$\_$J003147$-$423523 &  00 31 47.7 & $-$42 35 23.2 & 1.5 & 1.5  &  0.163  &   6.46 &   0.970  &  0.289 & S1$\_$1 &           \\
  ELAISC15$\_$J003214$-$425339 &  00 32 14.4 & $-$42 53 39.6 & 1.4 & 1.5  &  0.173  &   6.87 &   0.944  &  0.281 & S1$\_$1 & star      \\
  ELAISC15$\_$J003218$-$430542 &  00 32 18.3 & $-$43 05 42.3 & 1.8 & 1.6  &  0.152  &   5.42 &   0.821  &  0.291 & S1$\_$1 &           \\
  ELAISC15$\_$J003221$-$430020 &  00 32 21.7 & $-$43 00 20.3 & 2.0 & 1.6  &  0.128  &   5.04 &   0.868  &  0.298 & S1$\_$1 &           \\
  ELAISC15$\_$J003225$-$430712 &  00 32 25.8 & $-$43 07 12.3 & 2.0 & 1.6  &  0.127  &   5.00 &   0.971  &  0.332 & S1$\_$1 &           \\
  ELAISC15$\_$J003228$-$430758 &  00 32 28.0 & $-$43 07 58.6 & 1.9 & 1.6  &  0.135  &   5.27 &   0.826  &  0.277 & S1$\_$1 &           \\
  ELAISC15$\_$J003233$-$431304 &  00 32 33.4 & $-$43 13 04.6 & 1.4 & 1.5  &  0.232  &   6.79 &   0.981  &  0.286 & S1$\_$1 &           \\
\hline 
\end{tabular}
\end{minipage}      
\end{table*}                    

\section{Conclusions}
\label{concl}
A new data reduction technique (the {\it Lari method}) has been successfully
applied to the 15 $\mu$m ISOCAM observations of one of the four main ELAIS 
fields (S1). This technique, based on the existence of two different time-scales 
in ISOCAM transients, was particularly efficient in overcoming the main problems
affecting the ISOCAM LW data and in detecting faint sources. Its application to
the southern ELAIS field has produced a catalogue of 462 sources, detected above
the 5 $\sigma$ threshold over an area of about 4 square degrees. The integrated
fluxes of these sources cover the range 0.5 - 100 mJy, filling the existing gap 
between the Deep ISOCAM Surveys and the Faint IRAS Survey.  
The completeness and photometry accuracy of our catalogue have been tested 
through accurate simulations performed at different flux levels.
The results of these simulations showed that our catalogue is highly reliable
and $> 98.5$\% complete at 3 mJy. The completeness, due either to the mapping
effects or to the data reduction method, then decreases at fainter fluxes.  
The positional accuracy, estimated with simulations, resulted to be
about 1 arcsec in both right ascension and declination for signal-to-noise
ratios $> 7$, while it increases to $\sim 2$ and 1.6 at signal-to-noise
ratios close to the survey threshold (5), respectively for right ascension and 
declination. The photometric accuracy of our data reduction has also been tested 
using the stars of the field, comparing the measured fluxes with the ones
predicted by the relation calibrated on IRAS data by Aussel \& Alexander (2000).
Our fluxes resulted well consistent with the predicted ones over a large
range of fluxes, since these stars go from 0.85 mJy to 135 mJy in LW3.

In a forthcoming paper (Gruppioni, Lari, Pozzi et al. 2001) we will present the 
source counts obtained from this survey in the crucial uncovered flux range 
0.45 -- 100 mJy, dividing the Deep/Ultra-Deep ISOCAM Surveys from the fainter IRAS 
Surveys.

\section*{Acknowledgments}              
This work was supported by the EC TMR Network programme FMRX--CT96--0068.
CG acknowledges partial support by the Italian Space Agency (ASI) under
the contract ARS--98--119 and by the Italian Ministry for University and
Research (MURST) under grants COFIN98 and COFIN99. We would like to thank  
Gianni Zamorani for helpful suggestions and for a careful reading of the 
manuscript and David Elbaz for constructive refereeing, which improved the 
quality of the paper.

\newpage
\appendix
\section{Lari Model Description}

The process is governed by two differential equations, one for each charge reservoir,
of the form:

\begin{equation}
  \frac{dQ}{dt} = e I - a Q^2 \label{eq:1}
\end{equation}

\nin where $I$ is the incident flux of photons, $e$ is the 
efficiency of the process feeding the component and $a$ is a time constant which 
depends on the detector pixel-size. Note that $e$ and $a$ assume different values
for the two components: $a(${\bf breve}) $> a(${\bf lunga}) and $e(${\bf breve}) $> 
e(${\bf lunga}).

We have not attempted to model $I$ and $e$ for glitches. In principle, glitches could
be described by the physics of ionising particles.  However their effect strongly
depends on the nature and energy of the incident cosmic particle. For example, high 
energy incident cosmic rays could produce saturation on the detector, thus causing the 
parameter $e$ to depend on the values of $I$ and $Q$ . For simplicity, we 
have neglected this effect in our model, considering both $e$ and $a$ as constants.

Our model is completely conservative (no decay of the accumulated charges is considered, except
toward the contacts) and homogeneous (the charge reservoir involves all the detector 
parts that do not contribute to polarising the contacts). In fact, at stabilisation we
have $a Q^2 = e I$ and $S = I$ ($S$ being the signal), while generally 
$I = S + \Delta Q / \Delta t$.
The quantity $-a Q^2$ in the accumulated charge equation is exactly the same amount of charge
which that component feeds the contacts with.

Considering charges as fluxes in ADUs ($q = Q / \Delta t$), we have:

\begin{equation}
 \Delta t~ \frac{dq_1}{dt} = e_1 I - a_1 \Delta t^2~ q_1^2  \label{eq:2}
\end{equation}

\nin for the {\bf breve} component. Integrating over an observation integration time $\Delta t$
(with $I =$ constant):

\begin{equation}
\Delta q_1 = e_1 I - a_1  \Delta t^2  <q_1^2>  \label{eq:3}
\end{equation}

\nin and

\begin{equation}
\Delta t~ \frac{dq_2}{dt} = e_2 I - a_2 \Delta t^2~ q_2^2  \label{eq:4}
\end{equation}

\nin for the {\bf lunga} component, which integrated over an integation time becomes

\begin{equation}
\Delta q_2 = e_2 I - a_2 \Delta t^2  <q_2^2>  \label{eq:5}.
\end{equation}

\nin We then have

\begin{equation}
  S (obs. ~signal) = e_0 I + a_1 \Delta t^2 <q_1^2> + a_2 \Delta t^2 <q_2^2>  \label{eq:6}
\end{equation}

\nin where $e_0 + e_1 + e_2 = 1$

\nin The two differential equations are of {\it Riccati type}, which, for  
$I = constant$ have a general analytical solution:

\begin{equation}
  {q(t)} = A \times \frac{((A+q(0))~e^{bt} - (A-q(0)))} {((A+q(0))~ e^{bt} + (A-q(0))}  \label{eq:7}
\end{equation}

\nin where $q$ can be either $q_1$ or $q_2$. $A$ represents the asymptotic value for $q$ ($A = \frac{\sqrt{e I / a}}{\Delta t}$),
while  $b$ ($= 2 \sqrt{e I a}$) is the inverse of the time--scale at stabilisation. Note that 
the time--scale in this model 
is inversely proportional to the square root of the (constant) incident flux (at stabilisation 
only!), while the observed flux $S$ tends to $I$ as $t$ tends to infinity.

In our code we make use also of an approximate equation with finite difference values to obtain
valid solutions also for the general case of variable $I$:

\begin{equation}
(q(t + \Delta t) - q(t)) = e I(t) - a \Delta t^2 q(t + \Delta t) q(t) \label{eq:8}
\end{equation}

\nin where $q$ is either $q_1$ or $q_2$, $e$ is either $e_1$ or $e_2$, $a$ is either $e_1$ or $e_2$,
respectively for the {\bf breve} and for the {\bf lunga} components. 
$q(t)$ and $q(t+\Delta t)$ are the charges that are accumulated respectively at the beginning and at 
the end of and integration. $I(t)$ is the average intensity over the whole integration.   

The error we commit by use of this second-order approximation instead of the exact equation
is less than 1\%, so this can be considered a good approximation.
The same kind of approximation is used in other parts of our code when calculating derivatives. 

In equation \ref{eq:6} the observed flux is the flux subtracted by the dark current.
In dark observations `glitch' transients show as `faders' and `dippers', the
latter having time-scales larger than the integration time, but not infinite (as the model requires). 
Thus, both {\bf lunga} and {\bf breve} charge productions are fed also by the thermal component
of the dark current. This effect might be important when the photon flux is small.
It is very difficult to estimate this extra source of transient signal and, since we could not
find any documentation on ISOCAM thermal dark current measurements, we tried to estimate it
from the data. We have thus associated the thermal dark current to the minimum amount of
extra signal that is needed in order to keep the parameter $e_2$ below the value of the {\bf lunga}
fraction ($\simeq 0.1$, implying that {\it dippers} cannot exceed one-tenth of the sky background counts).

\nin By giving the name `offset' to the thermal feeding, equations \ref{eq:3}, \ref{eq:5} and 
\ref{eq:6} become:

\begin{equation}
\Delta q_1= e_1 (I + offset) - a_1 \Delta t^2 <q_1^2> \label{eq:9}
\end{equation}

\begin{equation}
\Delta q_2= e_1 (I + offset) - a_2 \Delta t^2 <q_2^2> \label{eq:10}
\end{equation}   

\begin{eqnarray}
 \nonumber \lefteqn{ S(obs.~ signal) + offset = } \\
& e_0 (I + offset) + a_1 \Delta  t^2  <q_1^2> + a_2 \Delta t^2 <q_2^2> \label{eq:11}
\end{eqnarray}

In practice, these two equations can be solved by successive iterations, provided that $q_{1/2} (0)$,
$e_{1/2}$ and $a_{1/2}$ are known. Estimates for these parameters, characterising each pixel,
can be obtained by minimising the $\chi^2$ estimator, under the condition of having a constant 
incident flux at each raster position.
In this model, all the past history of each pixel is contained in the intial charge, as long
as there is no other source of electrons (eg from the surrounding pixels or from longer
time relaxation processes inside the pixel).


\begin{thebibliography}{}
\bibitem[Akin(1970)]{akin70} Akin H. and Colton R.R., 1970, {\em Statistical Methods}, Fifth
Barnes \& Nobles Books Edition
\bibitem[Alexander(2000)]{davo00} Alexander A. \& Aussel H., in D. Lemke, M. Stickel
and K. Wilke eds., ISO Surveys 
of a Dusty Universe, Springer Lecture Notes of Physics Series, p. 113
\bibitem[Aussel(1999)]{auss99} Aussel H., Cesarsky C.J., Elbaz D. and Starck J.-L., 
1999, A\&A, 342, 313
\bibitem[Cesarsky(1996)]{ces96} Cesarsky C.J., Abergel A., Agn\`esel P. et al., 1996, A\&A, 315, L32
\bibitem[Elbaz(1999)]{elb99} Elbaz D., Cesarsky C. J., Fadda D., Aussel H., D\'esert
F.X., Franceschini A., Flores H., Harwit M., Puget J.-L., Starck J.-L., Clements 
D.L., Danese L., Koo D.C. and Mandolesi R., 1999, A\&A, 351, L37
\bibitem[Flores(1999)]{flor99} Flores H., Hammer F., Thuan T.X.,
C\'esarsky C., Desert F.X., Omont A., Lilly S.J., Eales S., Crampton D. and
Le Fèvre O., 1999, ApJ, 517, 148
\bibitem[Gruppioni(1999)]{grup99} Gruppioni C., Ciliegi P., Rowan--Robinson M., Cram L.,
Hopkins A., Cesarsky C., Danese L., Franceschini A., Genzel R., Lawrence A., 
Lemke D., McMahon R.G., Miley G., Oliver S., Puget J.-L. and Rocca-Volmerange B.,
1999, MNRAS, 305, 297
\bibitem[Gruppioni(2000)]{grup00} Gruppioni C., Lari C., Pozzi F., Zamorani G. and Franceschini A.,
2001, MNRAS, submitted 
\bibitem[Hog(2000)]{hog00} H\"og E., Fabricius C., Makarov V.V., Bastian U., Schwekendiek P., Wicenec A.,
Urban S., Corbin T. and Wycoff G., 2000, A\&A, 357, 367
\bibitem[Kessler(1996)]{kess96} Kessler M., Steinz J., Anderegg M. et al., 1996, A\&A, 315, L27
\bibitem[Lejeune(1998)]{lej98} Lejeune T. and Cuisinier F. and Buser R., 1998, A\&ASS, 130, 65
\bibitem[Monet(1998)]{mon98} Monet D.G., 1998, AAS Meeting, 193, p. 120.03 
\bibitem[Okumura(1998)]{oku98} Okumura K., 1998, {\em ISOCAM PSF Report}, available at 
http://www.iso.vilspa.esa.es/users/expl$\_$lib/CAM$\_$list.html
\bibitem[Okumura(2000)]{oku00} Okumura K., 2000, {\em ISOCAM Field Distortion Report}, available at 
http://www.iso.vilspa.esa.es/users/expl$\_$lib/CAM$\_$list.html
\bibitem[Oliver(2000)]{seb00} Oliver S., Rowan--Robinson M., Alexander D.M. et al., 2000, MNRAS, 316, 749 
\bibitem[Omont(1999)]{omont99} Omont A., Ganesh S., Alard C., Blommaert J.A.D., Caillaud B. et al., 1999,
A\&A, 348, 755
\bibitem[Schlegel(1998)]{schl98} Schlegel D.J., Finkbeiner D.P. and Davis M., 1998, ApJ, 500, 525 
\bibitem[Serjeant(2000)]{steve00} Serjeant S., Oliver S., Rowan--Robinson M. et al., 2000, MNRAS, in press
\bibitem[Starck(1999)]{starck99} Starck J.-L., Aussel H., Elbaz D., Fadda D. and Cesarsky C., 1999,
A\&AS, 138,365
\end{thebibliography}
\end{document}